\documentclass[12pt]{article}

\usepackage[table,xcdraw]{xcolor}
\usepackage{lscape}
\usepackage{hyperref,bbm}
\usepackage{amsfonts,amsmath,amssymb,amsthm}
\usepackage{amsbsy,amsthm,mathtools}
\usepackage{enumitem}  
\usepackage{verbatim} 
\usepackage{color}
\usepackage{graphicx}
\usepackage{multirow}
\usepackage[table,xcdraw]{xcolor}
\usepackage{natbib}
\usepackage{booktabs}
\usepackage{multirow}

\bibpunct{(}{)}{;}{a}{,}{,}
\newcommand{\biblist}{
\bibliographystyle{apalike}
\bibliography{postsel-references}
}
\usepackage{subfiles}
\usepackage{xr}
\usepackage{color}

\definecolor{dgreen}{rgb}{0.,0.6,0.}

\newtheorem{theorem}{Theorem}[section]
\newtheorem{corollary}{Corollary}[section]

\newtheorem{proposition}{Proposition}[section]
\newtheorem{assumption}{Assumption}[section]
\newtheorem{lemma}[theorem]{Lemma}

\newtheorem*{claim*}{Claim}
\newtheorem*{remark*}{Remark}
\newtheorem*{notation*}{Notation}

\makeatletter
\newcommand*{\defeq}{\mathrel{\rlap{%
                     \raisebox{0.3ex}{$\m@th\cdot$}}%
                     \raisebox{-0.3ex}{$\m@th\cdot$}}%
                     =}
\makeatother

\newcommand{\dd}[2]{\frac{\partial#1}{\partial#2}}

\newcommand{\mc}[1]{\mathcal{#1}}
\newcommand{\mb}[1]{\mathbbm{#1}}

\def\spacingset#1{\renewcommand{\baselinestretch}%
{#1}\small\normalsize} \spacingset{1.}

\begin{document}
\begin{center}
{\Large\textbf{Selective inference using randomized group lasso estimators for general models}}
\\
\bigskip

\textbf{Yiling Huang$^{1}$, Sarah Pirenne$^{2}$, Snigdha Panigrahi$^{1}$ and Gerda Claeskens$^{2}$} \\
\bigskip

$^1$ Department of Statistics, University of Michigan, Ann Arbor,  MI 48109-1107, U.S. \\
$^2$ Orstat and Leuven Statistics Research Center, KU Leuven, B-3000 Leuven, Belgium \\
\bigskip

yilingh@umich.edu; sarah.pirenne@kuleuven.be; psnigdha@umich.edu; gerda.claeskens@kuleuven.be

\end{center}

\begin{abstract} 
Selective inference methods are developed for group lasso estimators for use with a wide class of distributions and loss functions. 
The method includes the use of exponential family distributions, as well as quasi-likelihood modeling for overdispersed count data, for example, and allows for categorical or grouped covariates as well as continuous covariates.
A randomized group-regularized optimization problem is studied.
The added randomization allows us to construct a post-selection likelihood which we show to be adequate for selective inference when conditioning on the event of the selection of the grouped covariates. This likelihood also provides a selective point estimator, accounting for the selection by the group lasso. Confidence regions for the regression parameters in the selected model take the form of Wald-type regions and are shown to have bounded volume. The selective inference method for grouped lasso is illustrated on data from the national health and nutrition examination survey while simulations showcase its behaviour and favorable comparison with other methods.
\medskip

\end{abstract}

\spacingset{1.7} 

\section{Introduction}
Model selection, whether obtained by means of information criteria or via regularization methods adds uncertainty to the decision process.
This problem has been recognized for a long time, starting with unregularized estimation and selection \citep{HurvichTsai1990,Draper1995,BucklandBurnhamAugustin1997,HjortClaeskens2003}, and later with regularized estimation, such as the lasso \citep{LeeSunSunTaylor2016}.

Our motivation for this work stems from two key factors: firstly, by the need for inference after regularized estimation with high-dimensional datasets that contain \textit{grouped} covariates, and secondly, by the need for selective inference methods for models other than Gaussian.
Examples include datasets with categorical variables such as the National Health and Nutrition Examination Survey (NHANES) \citep{NHANES2017_2020} data, which is analyzed later in our paper.
In this dataset, the response is binary and variables like ethnicity and marital status have five and three categories, respectively.
Other examples with grouped covariates include non- and semiparametric models, where a function of a covariate is expanded using a spline or wavelet basis.
The collection of basis functions used to represent such a function is treated as a group.
In other instances, groups may be context-specific, where variables related to a particular characteristic or variables arising from common factors, are grouped.

For model selection purposes, it makes most sense to select such groups of variables together by using $\ell_2$-norm regularization, which results in the group lasso estimator \citep{YuanLin2006, meier2008group, roth2008group}.
However, the problem of inferring from non-normal data after selecting covariate groups is still not well addressed.
For instance, while the selective inference approach of \citet{BerkBrownBujaZhangZhao2013, BachocPreinerstorferSteinberger2016} offer a marginal, selection-agnostic viewpoint in Gaussian and logistic linear models respectively, confidence intervals are typically wide, and computations may not be convenient.
An alternative approach conditions on the event of selection to account for the randomness involved in the selection process.
For the lasso estimator in Gaussian linear models, \citet{LeeSunSunTaylor2016} use the polyhedral shape of the selection region to construct selective inference.
\citet{TaylorTibshirani2018} apply similar ideas to generalized linear models and Cox regression models.
Although this approach includes an additional conditioning on the signs of the estimated coefficients for computational convenience, like the marginal approach, it frequently suffers from a power loss.

Recently, there have been efforts to improve the power of selective inference through conditioning.
\citet{DuyVoNguyenLe2022} propose a computationally efficient approach based on parametric programming, which eliminates the need for additional conditioning on the signs of the nonzero estimated coefficients.
Alternatively, \citet{PanigrahiTaylor2022} and \citet{panigrahi2022exact} adopt the randomized framework in \cite{TianTaylor2018} to remedy the loss in power.
They construct approximate and exact confidence intervals, respectively, when working with normal data.
These methods rely on the polyhedral shape of the selection region, similar to the previous conditional methods.
But the presence of the $\ell_2$-norm in the regularization makes it difficult to apply the existing methodology or its direct modifications to group lasso estimators where there is no easy characterization of the selection event.

In this paper, we provide frequentist inferences using an asymptotic post-selection likelihood for group lasso estimators.
To achieve this, we incorporate an extra Gaussian random variable into the objective function to be minimized.
We use the term ``randomization" to refer to this externally added random variable. 
Our method builds on the work by \cite{PanigrahiMacDonaldKessler2021} which employs extra randomization to obtain Bayesian inferences after selecting groups of variables.
However, our current method is more versatile and has a broader scope compared to this previous work, which was limited to a fixed $X$ regression setting with a Gaussian response variable.
As will be evident in the paper, our novel contributions include the following:
\begin{enumerate}[nolistsep,leftmargin=3ex]
\item Through the encompassing concept of M-estimation, our method can be used for models beyond Gaussian and estimation methods beyond least squares.
The asymptotic post-selection likelihood in our method is developed to be applicable to likelihood models, including the class of generalized linear models, and covers more general forms of estimation, such as quasi-likelihood estimation, which can accommodate overdispersion.
\item Our post-selection confidence intervals assume the same form as the classical Wald intervals, and as a by-product, we also have a point estimator.
\item Our method for selective inference involves solving an easy optimization problem, which is convex and $|\mathcal{G}_E|$-dimensional, where $|\mathcal{G}_E|$ is the number of selected groups.
This method can be applied to ungrouped covariates as well, for group sizes equal to one.
Unlike the Bayesian approach taken by \cite{PanigrahiMacDonaldKessler2021}, our method is more efficient and does not rely on sampling.
\end{enumerate}

Randomization is utilized in our current paper to achieve convergence of a post-selection likelihood to an asymptotic function, enabling selective inference.
Several recent papers have demonstrated other benefits of incorporating randomization.
For example, \citet{MinSeunghyun2021Ccsa} develop confidence sets using a randomized lasso estimator after model selection.
\citet{RacinesYoung2022} link randomized response to data splitting.
\citet{LeinerDuanWassermanRamdas2021} employ a data fission approach, while \citet{dharamshi2023generalized} utilize a data thinning procedure, both of which involve randomization.
\cite{zhang2023framework} obtain the limiting distribution of estimators based on randomized sketching or projections.
Finally, \citet{DaiLinXingLiu2023} propose a Gaussian mirror that adds a Gaussian random variable to the covariates to control the false discovery rate.

Our method can be also seen related to the randomized method of data carving \citep{FithianSunTaylor2014}, which was described for linear models.
\citet{PanigrahiSnigdha2021Imfp, Panigrahi2022integrative} introduce Bayesian techniques to implement data carving and account for the selection process by marginalizing over the added randomization.
Pivots based on data carving are shown to yield valid asymptotic post-selection inference in \cite{Panigrahi2018carving}.
\citet{SchultheissRenauxBuehlmann2021, Liu2023selective} aggregate data carving over different splits to improve the selection stability in high-dimensional generalized linear models.
However, multicarving has not yet been developed for parameters selected by the group lasso.

The rest of the paper is organized as follows.
The estimation framework and notations are presented in Section \ref{sec:randomization}, which is followed by a first example illustrating the advantages of randomization.
The main methodological results are contained in Section~\ref{sec:method}, while Section~\ref{sec:inference} explains our inference methods.
Examples of where this method applies are given in Section~\ref{sec:examples}. 
Our numerical work is summarized in Section~\ref{sec:simul}, the analysis of the NHANES data example can be found in Section~\ref{sec:data} and Section~\ref{sec:disc} concludes.
All proofs are collected in the Appendix.

\section{Notation and estimation framework}\label{sec:randomization}

\subsection{Randomized selection}
For a sample size $n$ we denote the vector of independent responses $Y=(Y_1,\ldots,Y_n)^\top$. The design matrix of covariates is $X\in\mathbbm{R}^{n\times p}$, with rows denoted by $X_i=(x_{i1},\ldots,x_{ip})$, for $i=1,\ldots,n$, which are vectors of length $p$.
We assume a potentially large but fixed number of parameters $p$. 

We suppose that the collection of covariates can be partitioned into disjoint groups, indexed by the known set $\mc{G}$ which is a partition of $\{1,\ldots,p\}$.
We denote by $X_g\in \mathbbm{R}^{n\times |g|}$ the submatrix of $X$ corresponding to group $g\in \mc{G}$, and let $\beta_g$ denote the accompanying vector of regression parameters.
All groups together have $p$ elements, and the full vector $\beta$ is $p$-dimensional. In case each group consists of a singleton, the covariates are not grouped.
For grouped data, variable selection takes place via a group lasso estimator \citep{YuanLin2006} which has the advantageous property that it does not separate the elements of a group during the selection.

Let the added random variable $\omega_n$ be such that $\sqrt{n}\omega_n \sim N_p(0,\Omega)$, where $\Omega$ is a user-specified covariance matrix.
We solve the following randomized regularized optimization problem
\begin{eqnarray} \label{eq:opt.problem}
\widehat\beta_n^{(\Lambda)} \in \underset{\beta\in\mathbbm{R}^p}{\text{argmin}}\left\{ \frac{1}{\sqrt{n}} \ell (X\beta;Y) + \sum_{g\in\mathcal{G}} \lambda_g\|\beta_g\|_2 -\sqrt{n}\omega_n^\top\beta\right\}
\end{eqnarray}
where
$$\ell(X\beta;Y)= \sum_{i=1}^n \rho(x_i^\top \beta; y_i)$$
is a loss function, convex in $\beta$, and twice differentiable, and $\lambda_g\in \mathbbm{R}^+$, for ${g\in\mc{G}}$, are the tuning parameters.
Later in this section, we provide a first example to illustrate the benefits of our Gaussian randomization scheme and the choice of $\Omega$.
Examples of loss functions are given in Section~\ref{sec:examples}, also including extensions that can be dealt with in a similar way.


We use $[p]$ as shorthand notation for the set $\{1,\ldots,p\}$.
For $\theta=(\theta_1,\ldots,\theta_n)^\top$, let
\begin{align*}
	\nabla \ell(\theta;Y) =
	\begin{bmatrix}
		\sum_{i=1}^n \dd{}{\theta_j}\rho(\theta_i; Y_i)
	\end{bmatrix}_{j\in[n]} =
	\begin{bmatrix}
		\dd{}{\theta_j}\rho(\theta_j; Y_j)
	\end{bmatrix}_{j\in[n]} \in \mb{R}^n
\end{align*}
denote the gradient of $\ell(\cdot;Y)$.
Define the $n\times n$ matrix of second order derivatives
  $\nabla^2 \ell(\theta; Y) = \text{diag}\{\frac{\partial^2}{\partial\vartheta^2}\rho(\vartheta;Y_i)
  \big|_{\vartheta=\theta_i}; i\in[n] \}$.
We denote the set of indices of the active (nonzero) entries of the  solution to the optimization problem in \eqref{eq:opt.problem} by
$$\widehat{E} = \left\{j\in[p]: \widehat\beta_{n,j}^{(\Lambda)} \neq 0\right\}.$$
Let $E$ be the realized value for $\widehat{E}$ with our specific data $(Y, X)$ and denote its complement in $[p]$ by $E'$.

For $a,b\in\mathbbm{N}$, We denote by $0_{a,b}$ a matrix of dimension $a\times b$ with all zero elements, $0_a$ is a vector of length $a$ consisting of zeroes only, $1_a$ is an $a$-vector consisting of ones, and $I_{a}$ is the $a\times a$ identity matrix. For a partitioned $a$-vector $v=(v_g,g\in\mc{G})^\top$ with $\mc{G}$ a partitioning of $[a]$ and $|\mc{G}|$ denoting the number of elements in $\mc{G}$, we denote the $a\times |\mc{G}|$ matrix $\text{bd}(v)=\text{block diag}(v_g;g\in\mc{G})$ the block-diagonal matrix of which the $k$th block, $k\in \{1,\ldots,\mc{G}\}$, consists of the $k$th subvector of $v$, according to the partitioning $\mc{G}$.

After selection, our target for inference is obtained by solving the population M-estimation problem with the selected predictors as
 \begin{equation}
 	b_{n,E}^* = \underset{b\in\mathbbm{R}^{|E|}}{\text{argmin}}\;  \frac{1}{\sqrt{n}}\mathbb{E}\left[ \ell(X_E b; Y)\right].
	\label{eqn:targetM}
 \end{equation}
Here, and elsewhere, all expectations are with respect to the unknown true distribution of the data $(X,Y)$. When defining this target, note that we do not make any assumptions about the quality of selection. We let $b_n^* = (b_{n,E}^{*\top}, 0_{|E'|}^\top)^\top$
 be the $p$-dimensional vector, containing $|E'|$ zeros.
  Additionally, define the
    $p\times p$ matrices
  \allowdisplaybreaks
	\begin{eqnarray}
	H &=& \frac{1}{n}\mathbb{E}\left[X^\top\nabla^2 \ell(X_E b_{n,E}^*; Y) X\right]
=   \begin{bmatrix}
	H_{E,E} & H_{E,E'} \\
	H_{E',E} & H_{E', E'}
	\end{bmatrix} =
	\begin{bmatrix} H_{E} & H_{E'}\end{bmatrix},
\label{eq:defH}\\
	K &= & \textup{Var}\left(
	\frac{1}{\sqrt{n}}X^{\top}\nabla \ell(X_E b_{n,E}^*; Y)
	 \right)
	=
	\begin{bmatrix}
	K_{E,E} & K_{E,E'} \\
	K_{E',E} & K_{E', E'}
	\end{bmatrix}
	=
	\begin{bmatrix} K_{E} & K_{E'}\end{bmatrix},
\nonumber
      \end{eqnarray}
based on the moments of the gradient and the Hessian of our loss function.

Denote by $\widehat{\beta}_{n,E}$ the corresponding $E$-restricted M-estimator, i.e., we have
 \begin{align*}
 	 \frac{1}{\sqrt{n}}X_E^\top \nabla \ell(X_E \widehat{\beta}_{n,E}; Y)  =0_{|E|}.
 \end{align*}
Note that we would make inferences using the naive Gaussian distribution of the M-estimator only when $E$ is fixed beforehand without using any data.

In our problem, the set $E$ is data dependent and hypotheses about parameters are only stated after one observes which coefficients are effectively selected.
Therefore, valid inference necessitates accounting for the effects of selection, which we achieve through the use of specific conditional distributions.
Before we derive an asymptotic conditional likelihood in Section \ref{sec:method}, we provide a first data example to build intuition and draw connections with the widely used data splitting.

\subsection{A first example}

We simulate a toy example using datasets containing $n$ independent and identically distributed instances of a binary response variable $Y$, along with five categorical predictor variables, each having five levels.
Among these predictor variables, two are related to the response variable through a logistic model with non-zero coefficients, while the remaining three have zero coefficients in this model. That is, we generate
\begin{align*}
    Y_i \sim Bernoulli\left(\frac{e^{X_i^\top \beta}}{1+e^{X_i^\top \beta}}\right),
\end{align*}
where $X_i = \left[
    \mathbf{x}_{i,1} \ \mathbf{x}_{i,2} \ \cdots \ \mathbf{x}_{i,5}\right]^\top$, and $\beta = [\beta_1 \ \beta_2 \ \cdots \beta_5]^ \top$, with $\beta_1, \beta_2 \neq \mathbf{0}$, and $\beta_3, \beta_4, \beta_5 = \mathbf{0}$.
We employ a logistic loss function, utilizing the binary response variable and one-hot encoded categorical predictors, to solve equation \eqref{eq:opt.problem}.    
%

A simple way to get correct inference after selection, whether or not by regularization, is the randomized data splitting.
This involves using a random subset of the samples for variable selection and the other independent subset for performing inference about the coefficients in the selected model.
In this toy example, we used $70\%$ of our data samples to select the relevant groups, and then used the remaining held out samples to construct inferences in the selected logistic regression model.

Of course, one might be tempted to use naive inference, selecting a logistic model and reusing the same data without considering the impact of selection.
In Table \ref{tab:covtable}, we examine coverage rates of $90\%$ confidence intervals for data splitting and naive inference from $500$ simulations.
Based on our findings, it is evident that data splitting provides valid inferences as we do not use the same sample for inference and selection.
Whereas naive inference, which is highlighted in red in the left part of the table, has a much lower coverage rate than the desired level of $90\%$, despite having more samples than predictors.

\begin{table}[h]
\centering
\setlength\extrarowheight{-5pt}
\resizebox{\linewidth}{!}{%
\begin{tabular}{@{}cccc|cc|cc@{}}
\toprule
&\multicolumn{3}{c}{Mean coverage rate} & 
\multicolumn{2}{c}{F1-scores} & \multicolumn{2}{c}{Mean interval lengths} \\
$n$   & Naive & Data Splitting  & Post-GL        
& Data Splitting & Post-GL & Data Splitting & Post-GL    \\ \midrule
200 & {\color[HTML]{FE0000} 0.760} & 0.890   & 0.894            & 0.386   & 0.397 &  {\color[HTML]{FE0000} 60.672} & 12.502             \\
350 & {\color[HTML]{FE0000} 0.755} & 0.888   & 0.880              & 0.427   & 0.428 & {\color[HTML]{FE0000} 14.253} & 12.626  \\
500 & {\color[HTML]{FE0000} 0.775} & 0.890   & 0.882
& 0.404   & 0.420 & {\color[HTML]{FE0000} 13.687} & 12.672 \\ \bottomrule
\end{tabular}
}
\caption{
\small{Results  under varying sample size $(n)$ with nominal 90\% confidence: mean coverage rates, F1-scores -- see \eqref{eq:def-F1score} for the definition, and mean interval lengths.}}
\label{tab:covtable}
\end{table}


Obviously, inferring with only a subset of the data with data splitting reduces the power as compared to results obtained for the full dataset.
This motivates our proposed method with extra randomization, which we call ``Post-GL", short for post-group lasso inference.
The use of a conditional distribution in our method allows for more powerful inferences without wasting any data, unlike data splitting.

To implement our method, a randomization variable $\omega_n$ is drawn from $N_p(0,\Omega)$, where $\Omega=f H$ where $H$ is defined in \eqref{eq:defH} and $f$ is a prespecified number between $(0,1)$.
This choice of $\Omega$ is made to roughly match the amount of information used for selection in the benchmark method of data splitting.
In particular, selecting $f = \dfrac{1-r}{r}$ leads to asymptotically equivalent selection as data splitting with $r$-fraction of the data samples employed for selection.

Deferring a formal guidance in this direction to Lemma \ref{S-lem:samplesplitting} in the Appendix, for now, we turn to empirical evidence to confirm this equivalence.
In the same setup as before, we compare the accuracy of model selection produced by the two randomized methods, data splitting and ``Post-GL" in the middle part of Table \ref{tab:covtable}.
This accuracy is measured by comparing the recovered support set from selected models to the true support set, using F1 scores,  see \eqref{eq:def-F1score} for the definition.
Note that the models selected by data splitting and ``Post-GL" roughly match in quality across all sample sizes $n$.
Furthermore, as seen in the left part of Table \ref{tab:covtable}, ``Post-GL" achieves the target $90\%$ coverage rate just as data splitting in all settings.


Finally, in the right-hand side of Table \ref{tab:covtable}, we compare the power of two randomized methods by calculating the average lengths of confidence intervals.
The comparison reveals that data splitting consistently constructs longer intervals than ``Post-GL" across all sample sizes.
This difference becomes more pronounced as the sample size gets smaller (i.e., $n=200$), where the data splitting intervals can become numerically unstable and produce much longer intervals than ours.
The bottomline is that ``Post-GL" generates shorter intervals than data splitting while achieving similar model selection quality.
Moreover, correcting for selection in our method, as described in the following sections, is as simple as solving a low-dimensional convex optimization problem.


\section{Methodology} \label{sec:method}

Following the notation introduced in Section \ref{sec:randomization}, we explain how we characterize the event of selection with randomization in Section \ref{sec:selection-characterization}. In Section~\ref{sec:asympt-postsel-likelihood}, we derive the likelihood, and provide an asymptotic justification for it.

\subsection{Randomized selection of groups: a simple characterization}\label{sec:selection-characterization}

We start from obtaining a simple characterization for a subset of the selection event
$\{\widehat{E}=E\}$.
Before we do so, we introduce some more notations.
Let $\mc{G}_E$ denote the set of selected groups.
For $g \in \mc{G}_E$, we can write
\begin{eqnarray*}
\widehat\beta_{n,g}^{(\Lambda)} = \widehat{\gamma}_{n,g} \widehat{u}_{n,g}, \mbox{ where }
	\widehat{\gamma}_{n,g} = \|\widehat\beta_{n,g}^{(\Lambda)}\|_2, \mbox{ and }
	\widehat{u}_{n,g} = {\widehat\beta_{n,g}^{(\Lambda)}}\big/{\|\widehat\beta_{n,g}^{(\Lambda)}\|_2}.
\end{eqnarray*}
Then, define
$ 
\widehat{\gamma}_n = (\widehat{\gamma}_{n,g}: g \in \mc{G}_E)^\top \in \mb{R}^{|\mc{G}_E|},\
\widehat{u}_n = (\widehat{u}_{n,g}: g \in \mc{G}_E)^\top,
$ 
 and  the $|E|\times|\mc{G}_E|$ matrix $\widehat{U}_n= \text{bd}(\widehat{u}_n)$.
In matrix notation, we have
$\widehat\beta_{n,E}^{(\Lambda)} = \widehat{U}_n \widehat{\gamma}_n$.
Further, we let
$
\widehat{z}_n = (\widehat{z}_{n,g}: g \in \mc{G}_{E'})^\top
$
denote the subgradient vector of the group lasso penalty in \eqref{eq:opt.problem} at the non-active (zero) groups.	

The following lemma shows that an appropriate subset of our selection event can be characterized simply through sign constraints on $\widehat{\gamma}_n$. 
Its proof is straightforward and the same property also holds without randomization.

Even though the regularized estimator $\widehat\beta^{(\Lambda)}_n$ does not appear explicitly in this result, its information is contained in the vectors $\widehat\gamma_n$, $\widehat u_n$ and $\widehat z_n$.

\begin{lemma}
	\label{lemma:event:equivalence}
For the randomized group lasso estimation as in \eqref{eq:opt.problem}, it holds that
	\begin{align*}
		\left\{\widehat{E}=E, \ \widehat{u}_n= u_n, \ \widehat{z}_n = z_n\right\} = \left\{\widehat{u}_n= u_n, \ \widehat{z}_n = z_n, \ \sqrt{n}\widehat{\gamma}_n \in \mb{R}^{|\mc{G}_E|}_+ \right\}.
	\end{align*}
\end{lemma}

Note that \cite{PanigrahiMacDonaldKessler2021} employed the above characterization to describe the selection event in the fixed $X$ regression setting. 
However, unlike this previous work, we use randomization in our method for a dual purpose. 
Apart from obtaining a description of the selection event, we utilize randomization to relate the group lasso estimators $\widehat{\gamma}_n$, $\widehat{u}_n$, and $\widehat{z}_n$ with the $E$-restricted M-estimator $\widehat{\beta}_{n,E}$ and characterize their post-selection asymptotic distribution, which is a nontrivial task.

\subsection{Asymptotic post-selection likelihood} \label{sec:asympt-postsel-likelihood}

We base selective inference on the conditional distribution of
$
\sqrt{n}  \widehat{\beta}_{n,E}
$
given the event in Lemma \ref{lemma:event:equivalence}.
The resulting likelihood, however, is not a function of $b_{n,E}^*$ alone, but involves nuisance parameters which are denoted by ${b}_{n,E}^\perp$.
We are able to eliminate these nuisance parameters in the limit, and subsequently obtain an asymptotic likelihood function in $b_{n,E}^*$ by conditioning further on the statistic
\begin{eqnarray}
\label{eq:def-betaperp}
\widehat{\beta}_{n,E}^{\perp}= \frac{1}{n}X_{E'}^{\top}
	\nabla \ell(X_E\widehat\beta_{n,E}; Y)- A_E\widehat{\beta}_{n,E} \in \mb{R}^{|E'|},
\end{eqnarray}
where $A_E = H_{E',E} -K_{E',E}K_{E,E}^{-1}H_{E,E}$.
Letting
$ 
	V_{n,E'} = \mathbb{E}\big[ \frac{1}{n}X_{E'}^{\top}\nabla \ell(X_E b_{n,E}^*; Y) \big],
$ 
we define the vector of nuisance parameters
${b}_{n,E}^\perp = V_{nE'}-A_Eb_{n,E}^*$.

Formally, to provide selective inference, we consider the distribution of
\begin{equation}
\sqrt{n} \big( \widehat{\beta}_{n,E}^\top, \widehat{\gamma}_n^\top \big)^\top \Big\lvert \{\widehat{\beta}^{\perp}_{n,E}=  \beta^{\perp}_{n,E}, \widehat{u}_n= u_n, \ \widehat{z}_n = z_n\}
\label{mar: distn}
\end{equation}
when truncated to the event $\{\sqrt{n} \widehat{\gamma}_n \in  \mb{R}^{|\mc{G}_E|}_+\}$.
Denote by
$
L_{n, u_n, z_n, \beta^{\perp}_{n,E}}\big({b}^*_{n,E}, {b}_{n,E}^\perp; \widehat{\beta}_{n,E}, \widehat{\gamma}_n\big)
$
the marginal likelihood based on the distribution in \eqref{mar: distn}.
Truncating this likelihood to the event $\big\{\sqrt{n} \widehat{\gamma}_n \in  \mb{R}^{|\mc{G}_E|}_+\big\}$ gives rise to the following conditional likelihood:
\begin{equation*}
\begin{aligned}
& \dfrac{L_{n, u_n, z_n, \beta^{\perp}_{n,E}}\left({b}^*_{n,E}, {b}_{n,E}^\perp; \widehat{\beta}_{n,E}, \widehat{\gamma}_n\right) }{\int L_{n, u_n, z_n, \beta^{\perp}_{n,E}}\left({b}^*_{n,E}, {b}_{n,E}^\perp; \beta_{n,E}, \gamma_n\right) 1_{\mb{R}^{|\mc{G}_E|}_+}(\sqrt{n}\gamma_n)d\beta_{n,E}   d\gamma_n} 1_{\mb{R}^{|\mc{G}_E|}_+}(\sqrt{n}\widehat{\gamma}_n)\\
&=\dfrac{L_{n, u_n, z_n, \beta^{\perp}_{n,E}}\left({b}^*_{n,E}, {b}_{n,E}^\perp; \widehat{\beta}_{n,E},  \widehat{\gamma}_n\right)}{\mathbb{P}\left[\sqrt{n} \widehat{\gamma}_n \in  \mb{R}^{|\mc{G}_E|}_+ \;  \Big\lvert \; \widehat{\beta}^{\perp}_{n,E}=  \beta^{\perp}_{n,E}, \widehat{u}_n= u_n, \ \widehat{z}_n = z_n\right]} 1_{\mb{R}^{|\mc{G}_E|}_+}(\sqrt{n}\widehat{\gamma}_n).
\end{aligned}
\end{equation*}
The corresponding log-likelihood is given by
\begin{equation}
\log L_{n, u_n, z_n, \beta^{\perp}_{n,E}}\left({b}^*_{n,E}, {b}_{n,E}^\perp; \widehat{\beta}_{n,E}, \widehat{\gamma}_n\right) -\log \mathbb{P}\left[\sqrt{n}\widehat{\gamma}_n \in \mb{R}^{|\mc{G}_E|}_+  \;  \Big\lvert \; \widehat{\beta}^{\perp}_{n,E}=  \beta^{\perp}_{n,E}, \widehat{u}_n= u_n, \ \widehat{z}_n = z_n\right].
\label{cond:lik:n}
\end{equation}

To derive what we call our asymptotic post-selection likelihood, we apply two steps.
In Step 1, we substitute the marginal likelihood
$
L_{n, u_n, z_n, \beta^{\perp}_{n,E}}\big({b}^*_{n,E}, {b}_{n,E}^\perp; \widehat{\beta}_{n,E},  \widehat{\gamma}_n\big)
$
with its asymptotic counterpart which we derive in Proposition \ref{prop2: joint beta gamma}.
In Step 2, we obtain the large-deviation limit of
$
\log \mathbb{P}\big[\sqrt{n}\widehat{\gamma}_n \in \mb{R}^{|\mc{G}_E|}_+  \mid \widehat{\beta}^{\perp}_{n,E}=  \beta^{\perp}_{n,E}, \widehat{u}_n= u_n, \ \widehat{z}_n = z_n\big]
$
in Theorem \ref{Th:LDP}.

Details of both steps are provided below.

Following \citet[][Ch.~6]{White1994} we make the following assumption on the loss function.
\addtocounter{assumption}{-1}
\begin{assumption}
\label{assumption:basic}
The function $\rho(\theta;Y)$ has two continuous derivatives with respect to the parameter vector $\theta$ for which the expectations are finite.
The matrix $H=H(b_{n,E}^*)$ as defined in \eqref{eq:defH} is finite, positive definite and continuous in its argument uniformly in $n$.
\end{assumption}

\textbf{Step 1}. In Propositions \ref{prop: Taylor:series:rep} and \ref{prop: joint beta gamma}, we obtain a connection between the M-estimator and the biased group lasso estimators through randomization and use this to characterize their marginal asymptotic distribution.
By combining this result with the simple characterization of the selection event in Lemma \ref{lemma:event:equivalence}, we are able to derive an adjustment for selection, making it one of our key contributions in the paper.

Define
$\Lambda_E=\text{diag}(\{\lambda_gI_{|g|};g\in \mathcal{G}_E\})$ and
$\Lambda_{E'}=\text{diag}(\{\lambda_gI_{|g|};g\not\in \mathcal{G}_E\})$,
$A_{\mathcal{E}} = -K_EK_{E,E}^{-1}H_{E,E}$, $B_{\mathcal{E}} = H_E$, 
$C_{\mathcal{E}} = ( 0_{|E'|, |E|}, I_{|E'|})^\top$, $D_{\mathcal{E}} = (I_{|E|}, 0_{|E|, |E'|})^\top\Lambda_{E}$.
Further, define $\Sigma_E = H_{E,E}^{-1}K_{E,E}H_{E,E}^{-1}$, and $\Sigma^{\perp}_E=K_{E',E'} - K_{E',E} K_{E,E}^{-1} K_{E, E'}$.

\begin{proposition}
    \label{prop: Taylor:series:rep}
Under Assumption~\ref{assumption:basic},
for fixed values $b \in \mathbbm{R}^{|E|}$, $b^{\perp}\in \mathbbm{R}^{|E'|}$, define
\begin{align}
\Pi_{b, b^{\perp}}: \mathbbm{R}^{|\mathcal{G}_E|}\!\times\!\mathbbm{R}^{|E|}\!\times\! \mathbbm{R}^{|E'|}\to \mathbbm{R}^p: (g,u,z)\mapsto
 A_{\mathcal{E}} b + B_{\mathcal{E}}\text{bd}(u) g+ C_{\mathcal{E}} (\Lambda_{E'}z + b^{\perp}) + D_{\mathcal{E}}u.
	\label{CoV}
\end{align}
We have the following representation for the randomized group lasso estimator in \eqref{eq:opt.problem},
\begin{align*}
	\sqrt{n}\omega_n   &=  \Pi_{\sqrt{n}\widehat{\beta}_{n,E}, \sqrt{n}\widehat{\beta}^{\perp}_{n,E} }( \sqrt{n}\widehat\gamma_n, \widehat{u}_n,  \widehat{z}_n) + o_p(1).
\end{align*}
\end{proposition}
\noindent To simplify the notation we define the $p$-vector $\sqrt{n}\overline{\omega}_n$ $=\Pi_{\sqrt{n}\widehat{\beta}_{n,E}, \sqrt{n}\widehat{\beta}^{\perp}_{n,E} }( \sqrt{n}\widehat\gamma_n, \widehat{u}_n,  \widehat{z}_n)$.
\begin{corollary}
\label{lem:omega_bar_gaussian}
Under the assumptions of Proposition \ref{prop: Taylor:series:rep},
	$\sqrt{n}\overline{\omega}_n \overset{d}{\rightarrow} W\sim N_p(0, \Omega)$.
\end{corollary}

\begin{assumption}
\label{assumption:abscts}
There is a sample size $n_0$ such that for all $n\ge n_0$, the distribution of
$
\sqrt{n} \big((\widehat{\beta}_{n,E}-b_{n,E}^*)^\top,  (\widehat{\beta}_{n,E}^\perp-b_{n,E}^\perp)^\top, \overline\omega_n^\top \big)^\top
$
admits a Lebesgue density $p_n$.
\end{assumption}
This assumption is weak and avoids exceptional situations.
In Proposition~\ref{prop1: marg dist} we prove that the vector in Assumption~\ref{assumption:abscts} has a limiting Gaussian distribution.

\begin{proposition} \label{prop: joint beta gamma}\label{prop0:CoVdensity}
Under Assumptions~\ref{assumption:basic} and \ref{assumption:abscts},
the joint density function of
$$
\big( \sqrt{n} \widehat{\beta}_{n,E}^\top, \sqrt{n} (\widehat{\beta}_{n,E}^\perp)^\top, \sqrt{n} \widehat\gamma_n^\top, \widehat{u}_n^\top, \widehat{z}_n^\top \big)^\top
$$
evaluated at  the vector
$
\big( \sqrt{n}{\beta}_{n,E}^\top, \sqrt{n} ({\beta}_{n,E}^\perp)^\top, \sqrt{n} \gamma_n^\top, {u}_n^\top, {z}_n^\top \big)^\top
$ is given by
$$
|\det D_{\Pi}(\sqrt{n}{\gamma}_n, u_n, z_n)|p_n\big(\sqrt{n}(\beta_{n,E}-b_{n,E}^*), \sqrt{n}(\beta^\perp_{n,E}-b_{n,E}^\perp), \Pi_{\sqrt{n}\beta_{n,E}, \sqrt{n}\beta^\perp_{n,E}}(\sqrt{n}\gamma_n, u_n, z_n)\big),
$$
where
$D_{\Pi}(\cdot)$ is the Jacobian matrix of the map $\Pi_{b, b^{\perp}}(\cdot)$ defined according to \eqref{CoV}.
\end{proposition}

We turn to the asymptotic distribution of $\widehat{\beta}_{n,E}$, $\widehat{\beta}_{n,E}^{\perp}$, and $\overline{\omega}_n$ for a fixed set $E$.
\begin{proposition}
Suppose that $E$ is a fixed subset of $[p]$.
Under Assumption~\ref{assumption:basic}, we have that as $n$ tends to infinity,
$$
	\sqrt{n} \begin{pmatrix} \widehat{\beta}_{n,E} - {b}^*_{n,E} \\ \widehat{\beta}^{\perp}_{n,E} - {b}_{n,E}^\perp\\ \overline\omega_n \end{pmatrix}
 	\stackrel{d}{\to}
 	N_{2p}\left( \begin{pmatrix}0_{|E|} \\ 0_{|E'|} \\ 0_p\end{pmatrix}, \begin{bmatrix}\Sigma_E & 0 & 0 \\ 0 & \Sigma^{\perp}_E & 0 \\  0 & 0 & \Omega\end{bmatrix}\right).
$$

\label{prop1: marg dist}
\end{proposition}

Replacing $p_n$ with the limiting Gaussian density in Proposition \ref{prop1: marg dist} gives us the asymptotic counterpart for the marginal likelihood
$
L_{n, u_n, z_n, \beta^{\perp}_{n,E}}\big({b}^*_{n,E}, {b}_{n,E}^\perp; \widehat{\beta}_{n,E},  \widehat{\gamma}_n\big),
$
which we refer to as the `asymptotic' likelihood, despite its dependence on $n$.
Before we can state the asymptotic conditional distribution in Proposition~\ref{prop2: joint beta gamma}, we define the following matrices to simplify the expressions.
Denote
$\Gamma_n=\operatorname{diag}(\{\sqrt{n}\gamma_{n,g} I_{|g|-1}: g \in \mathcal{G}_E\})$,
\begin{align*}
\begin{gathered}
\overline{\Omega}=\left((B_{\mathcal{E}}{U}_n)^{\top} \Omega^{-1} B_{\mathcal{E}}{U}_n\right)^{-1},
\overline{A}=-\overline{\Omega} (B_{\mathcal{E}}{U}_n)^{\top} \Omega^{-1} A_{\mathcal{E}}, \\
\overline{b}=-\overline{\Omega} (B_{\mathcal{E}}{U}_n)^{\top} \Omega^{-1} (C_{\mathcal{E}} \Lambda_{E'} {z}_n+  D_{\mathcal{E}}  {u}_n  + C_{\mathcal{E}}\sqrt{n}{\beta}^{\perp}_{n,E}), \\
\overline{\Theta}=\left(\Sigma_E^{-1}-\overline{A}^{\top}\overline{\Omega}^{-1} \overline{A}+A_{\mathcal{E}}^{\top} \Omega^{-1} A_{\mathcal{E}}\right)^{-1},
\overline{R}=\overline{\Theta} \Sigma_E^{-1}, \\
\overline{s}=\overline{\Theta}\left(\overline{A}^{\top}\overline{\Omega}^{-1} \overline{b}-A_{\mathcal{E}}^{\top} \Omega^{-1} (C_{\mathcal{E}}  \Lambda_{E'} {z}_n+  D_{\mathcal{E}}  {u}_n  + C_{\mathcal{E}} \sqrt{n}{\beta}^{\perp}_{n,E})\right).
\end{gathered}
\end{align*}
For $g\in\mc{G}$, we define by $\overline{U}_{n,g}$ the orthogonal completion for $u_{n,g}$, which can be constructed via a Gram-Schmidt procedure.
Thus $\overline{U}_{n,g}$ is a matrix of dimension $|g|\times(|g|-1)$ of which the columns form an orthonormal basis for the tangent space
$
T_{u_{n,g}} \mathcal{S}^{|g|-1}=\left\{v\in \mb{R}^{|g|}: v^{\top} u_{n,g}=0\right\}
$
at $u_{n,g} \in \mathcal{S}^{|g|-1}$, the $(|g|-1)$-dimensional unit sphere which is embedded in $\mb{R}^{|g|}$.
For all groups combined this results in a matrix $\overline U_n=\text{block diag}(U_{n,g}; g\in \mc{G}_E)$ with dimension $|E|\times (|E|-|\mc{G}_E|)$.

\begin{proposition}
\label{prop2: joint beta gamma}
Assume the conditions in Proposition \ref{prop1: marg dist} and using the notation defined above,
let
$
J(\sqrt{n}\gamma_n, u_n)= \operatorname{det}\big(\Gamma_n+\overline{U}_n^{\top} H_{E,E} ^{-1} \Lambda \overline{U}_n\big).
$
The asymptotic distribution of
$$
\sqrt{n} (\widehat{\beta}_{n,E}^\top, \widehat{\gamma}_n^\top)^\top\Big\lvert\{ \widehat{\beta}^{\perp}_{n,E}=  \beta^{\perp}_{n,E}, \widehat{u}_n= u_n, \ \widehat{z}_n = z_n\}
$$
is represented by the following likelihood
\begin{equation} \label{eq:asymptlik}
d(b_{n,E}^*) J(\sqrt{n}\widehat{\gamma}_n, U_n) \phi(\sqrt{n}\widehat{\beta}_{n,E}; \overline{R}\sqrt{n} b^{*}_{n,E}+ \overline{s}, \overline{\Theta})  \phi(\sqrt{n}\widehat{\gamma}_n; \overline{A}\sqrt{n}\widehat{\beta}_{n,E}+ \overline{b}, \overline{\Omega}),
\end{equation}
\text{with}
$1/d(b_{n,E}^*)=\int
J(\sqrt{n}g, U_n) \phi(\sqrt{n}b; \overline{R}\sqrt{n} b^{*}_{n,E}+ \overline{s}, \overline{\Theta})  \phi(\sqrt{n}g; \overline{A}\sqrt{n}b+ \overline{b}, \overline{\Omega})db dg
$.
\end{proposition}

\textbf{Step 2}.
Our main result in Theorem \ref{Th:LDP} gives us a large deviation limit for the density related to $\sqrt{n}\widehat{\beta}_{n,E}$.
Before we state our assumptions, note from the proof of Proposition~\ref{prop1: marg dist} that
\begin{eqnarray} \label{eq:define Z}
\sqrt{n} \big( (\widehat{\beta}_{n,E} - {b}^*_{n,E})^\top, (\widehat{\beta}^{\perp}_{n,E} - {b}_{n,E}^\perp)^\top, \overline\omega_n^\top \big)
=T\sqrt{n}\overline{Z}_n + \tilde r_n,
\end{eqnarray}
where  $\tilde r_n= o_p(1)$, $\overline{Z}_n= \frac{1}{n} \sum_{i=1}^n Z_{i,n}$ is the mean of $n$ independent and identically distributed observations, and
$    T = \begin{bmatrix}
    	-H_{E,E}^{-1} & 0_{|E|, |E'|} & 0_{|E|, p} \\
    	H_{E',E}-A_E & I_{|E'|} & 0_{|E'|, p}\\
    	0_{p, |E|} & 0_{p, |E'|} & I_{p}
    \end{bmatrix}
$
is a fixed, invertible matrix.
We further observe that
	$T\sqrt{n}\overline{Z_n} \overset{d}{\rightarrow} N_{2p}(0_{2p}, \Sigma)$ where
	$\Sigma =\text{diag}(\Sigma_{E},\Sigma_{E}^\perp, \Omega)$.
The above representation is detailed out in the proof of Proposition \ref{prop1: marg dist}.

In order to apply a large deviation principle, we make the following assumptions.
\begin{assumption}[Moment condition and convergence of remainder]
\label{as:1}
Assume for the random variables defined in \eqref{eq:define Z} that
$\mathbb{E}\left[\exp(\lambda\|Z_{1,n}\|_2)\right]<\infty$
for some $\lambda\in \mb{R}^{+}$,
and that for a sequence with $a_n= o(n^{1/2})$, and $a_n\to \infty$ as $n\to \infty$, for any $\epsilon >0$
\begin{equation*}
\displaystyle\lim_{n\to \infty}  \frac{1}{a_n^{2}}\log \mathbb{P}\left[\frac{1}{a_n}\| \tilde r_n \|_2 > \epsilon \right] =- \infty.
 \end{equation*}
\end{assumption}

 \begin{assumption}[]
 \label{as:2}
Let $\mathcal{R}\subseteq \mb{R}^{2p}$ be a convex set.
For the random variables defined in \eqref{eq:define Z}, and for any sequence of random variables $C_n$ with bounded support and with $a_n= o(n^{1/2})$, and $a_n\to \infty$ as $n\to \infty$,
\begin{equation*}
\lim_{n\to \infty} \dfrac{1}{a_n^2} \left\{ \log \mathbb{P}\left[ \frac{\sqrt{n}}{a_n} \overline{Z}_n \in \mathcal{R}\right] - \log \mathbb{P}\left[ \frac{\sqrt{n}}{a_n} \overline{Z}_n + \frac{C_n}{a_n}\in \mathcal{R} \right]\right\}=0.
\end{equation*}
\end{assumption}

Due to the positivity constraint $\sqrt{n}\widehat{\gamma}_n \in \mb{R}^{|\mc{G}_E|}_+$, see Lemma~\ref{lemma:event:equivalence}, we work with a barrier function that replaces the non-differentiable indicator function which is $\infty$ for negative values and zero for strictly positive values.
In general, for any vector $v$, to incorporate the linear constraints $v_k>c$ for $k=1,2,\ldots, |v|$, we define $\text{Barr}(v)=\sum_{k=1}^{|v|} \log(1+1/(v_k-c))$.
\begin{theorem} \label{Th:LDP}
Denote $\mathcal{K}=[c, \infty)^{|\mathcal{G}_E|}$. Suppose $\sqrt{n}\big( (b_{n,E}^*)^\top, (b_{n,E}^\perp)^\top \big)^\top= a_n \big( (b_E)^\top, (b_E^\perp)^\top \big)^\top,$ where $a_n= o(n^{1/2})$, and $a_n\to \infty$ as $n\to \infty$.
	Let
	\begin{align*}
	O_n& = -\inf_{b', g'}
		\bigg\{ \frac{1}{2}\left(b' - \overline{R}b_E - \frac{1}{a_n}\overline{s}\right)^\top \overline{\Theta}^{-1}\left(b' - \overline{R}b_E - \frac{1}{a_n}\overline{s}\right)\\
		&\;\;\;\;\;+ \frac{1}{2}\left(g' - \overline{A}b' - \frac{1}{a_n}\overline{b}\right)^\top \overline{\Omega}^{-1}\left(g' - \overline{A}b' - \frac{1}{a_n}\overline{b}\right) - \frac{1}{a_n^2}\log J(a_ng'; U_n)+\frac{1}{a_n^2}{\text{\normalfont Barr}}_{\mathcal{K}}(a_n g')\bigg\},
        \end{align*}
    Under Assumptions \ref{as:1} and \ref{as:2}, we have
$$
\lim_{n\rightarrow\infty} \frac{1}{a_n^2} \log \mathbb{P}\bigg[ \sqrt{n}\widehat\gamma_{n}\in \mathcal{K}\ \bigg{|}\
		\widehat\beta_{n,E}^{\perp} = \beta_E^\perp, \widehat{u}_n = u_n, \widehat{z}_n = z_n		\bigg]  -O_n =0.$$
\end{theorem}
As a consequence of Theorem \ref{Th:LDP}, we can substitute the log-probability
$$
\log \mathbb{P}\left[\sqrt{n}\widehat{\gamma}_n \in \mathcal{K}  \;  \Big\lvert \; \widehat{\beta}^{\perp}_{n,E}=  \beta^{\perp}_{n,E}, \widehat{u}_n= u_n, \ \widehat{z}_n = z_n\right]
$$
 by the optimization
\begin{eqnarray} \label{opt:a}
a_n^2 O_n &= & -\inf_{b', g'}
		\bigg\{ \frac{1}{2}\left(a_nb' - \overline{R}\sqrt{n} b^*_{n,E} - \overline{s}\right)^\top\overline{\Theta}^{-1}\left(a_n b' - \overline{R}\sqrt{n}b^*_{n,E} - \overline{s}\right)\\
		&& + \frac{1}{2}\left(a_n g' - \overline{A}a_n b' - \overline{b}\right)^\top\overline{\Omega}^{-1}\left(a_n g' - \overline{A}a_n b' - \overline{b}\right) - \log J(a_ng'; U_n)+{\text{\normalfont Barr}}_{\mathcal{K}}(a_n g')\bigg\}.\nonumber
\end{eqnarray}
Of course, $a_n$ is not known in our problem.
But, simply substituting $a_nb' = \sqrt{n}b$, and $a_ng' =\sqrt{n} g$ in \eqref{opt:a},
completes Step 2, and gives us an expression for our asymptotic post-selection likelihood.
Ignoring constants, the logarithm of this likelihood agrees with
\begin{equation}
 \begin{aligned}
 	& \log \phi(\sqrt{n}\widehat{\beta}_{n,E}; \overline{R}\sqrt{n} b^{*}_{n,E}+ \overline{s}, \overline{\Theta})  + \inf_{b, g}
		\bigg\{ \frac{1}{2}\left(\sqrt{n} b - \overline{R}\sqrt{n} b^*_{n,E} - \overline{s}\right)^\top\overline{\Theta}^{-1}\left(\sqrt{n} b - \overline{R}\sqrt{n}b^*_{n,E} - \overline{s}\right)\\
		&\;+ \frac{1}{2}\left(\sqrt{n} g - \overline{A}\sqrt{n} b - \overline{b}\right)^\top\overline{\Omega}^{-1}\left(\sqrt{n} g - \overline{A}\sqrt{n} b - \overline{b}\right) - \log J(\sqrt{n}g; U_n)+{\text{\normalfont Barr}}_{\mathcal{K}}(\sqrt{n}g)\bigg\}.
 \end{aligned}
 \label{asymptotic:pslik}
\end{equation}
Finally, note that we are interested in
$$\mathbb{P}\bigg[ \sqrt{n}\widehat\gamma_{n}\in \mathbb{R}^{|\mathcal{G}_E|}_{+}\ \bigg{|}\
		\widehat\beta_{n,E}^{\perp} = b_E^\perp, \widehat{u}_n = u, \widehat{z}_n = z	\bigg].
$$
In practice, we may choose $c$ to be a small positive constant such that the above-stated probability is close to the probability in Theorem \ref{Th:LDP}.
If the Jacobian is bounded away from zero, one might take $c=0$, as we did in the simulations.

\section{Inference using the asymptotic post-selection likelihood}
\label{sec:inference}

We now work with the asymptotic post-selection log-likelihood in \eqref{asymptotic:pslik}. Its corresponding maximum likelihood estimator (MLE) is denoted by
$\widehat{b}_{n,E}^{mle}$, and the observed Fisher information matrix by $I_n(\widehat{b}_{n,E}^{mle})$.
In line with \citet{PanigrahiTaylor2022} for Gaussian regression with the lasso penalty, we show in Theorem~\ref{MLE:EstimatingEqn}
that the solution of a $|\mathcal{G}_E|$-dimensional optimization problem from \eqref{asymptotic:pslik} now gives us estimating equations for the MLE and
an expression for the observed Fisher information matrix.

\begin{theorem}
\label{MLE:EstimatingEqn}
Under assumptions \ref{assumption:basic}--\ref{as:2},
	consider solving the $|\mathcal{G}_E|$-dimensional optimization problem
$	g_n^*(\sqrt{n}\widehat{\beta}_{n,E}) = $
	\begin{equation}
 	\underset{g}{\arg\min}\big\{
	\frac{1}{2}(g - \overline A\sqrt{n}\widehat{\beta}_{n,E} - \overline b)^\top \overline\Omega^{-1} (g - \overline A\sqrt{n}\widehat{\beta}_{n,E} - \overline b)
	- \log J(g; U_n)
	+ {\text{\normalfont Barr}}_{\mathcal{K}}(g)\big\}.
\label{opt:esteqn}
\end{equation}
Then, the maximum likelihood estimator is obtained as
\begin{eqnarray*}
	\sqrt{n}\widehat{b}_{n,E}^{mle} = \overline{R}^{-1}\sqrt{n}\widehat{\beta}_{n,E} - \overline{R}^{-1}\overline{s} + \Sigma_{E} \overline A^\top \overline\Omega^{-1}\left(\overline A\sqrt{n}\widehat{\beta}_{n,E}+ \overline b- g_n^*(\sqrt{n}\widehat{\beta}_{n,E})\right).
\end{eqnarray*}
%
	Let the $|E|\times|E|$ matrix
$	
M = \overline \Theta ^{-1} + \overline A^\top\overline\Omega^{-1} \overline A
	- \overline A^\top\overline\Omega^{-1} \big[\overline\Omega^{-1} -\nabla^2 \log J(g_n^*(\sqrt{n}\widehat{\beta}_{n,E}); U_n)$ $+ \nabla^2{\text{\normalfont Barr}}_{\mathcal{K}}(g_n^*(\sqrt{n}\widehat{\beta}_{n,E}))\big] ^{-1}\overline\Omega^{-1} \overline A.
$
	The observed Fisher information matrix based on the asymptotic log-likelihood in \eqref{asymptotic:pslik} is equal to
$
		I_{n,\text{mle}} = n\Sigma^{-1}_{E}M^{-1}\Sigma^{-1}_{E}.
$
	\end{theorem}
This theorem provides the ingredients for conditional inference regarding $b^*_{n,E}$, by explicitly taking into account the selection of the set $E$ by the group lasso method.
For example, when we wish to construct a $1-\alpha$ confidence interval for the $j$th component of $b^*_{n,E}$, for $j\in\{1,\ldots,|E|\}$,
interval estimates for $b^*_{n,E}$, using our post-selection asymptotic likelihood, take the familiar Wald-type form centered around the selective MLE,
$$
\widehat{b}_{n,E}^{mle}[j]  \pm z_{1-\alpha/2}\sqrt{I^{-1}_{n,\text{mle}}[j,j]}.
$$
Likewise, we can construct confidence regions for $b^*_{n,E}$ (and similarly for a subvector thereof), that take the form
$
\{ b^*_{n,E} \in \mb{R}^{|E|}: \  (\widehat{b}_{n,E}^{mle}  - b^*_{n,E})^\top I_{n,\text{mle}}  (\widehat{b}_{n,E}^{mle}  - b^*_{n,E}) \leq \chi_{|E|, 1-\alpha}^2 \}.
$
For example, we might want a confidence region for the different levels of a categorical covariate that is selected by our randomized regularized problem.

A welcome consequence of the added randomization is that the confidence intervals using our method are bounded, which we formalize in the following lemma.
This is in contrast with the infinite expected length of the intervals for the (not randomized) selective inference results based on the polyhedral methods \citep[see][]{KivaranovicLeeb2021}.

\begin{lemma}
\label{lem:boundonvariance}
Under the previous assumptions, it holds that
\begin{eqnarray}
\max_{i,j\in\{1,\ldots,|E|\}}|(I^{-1}_{n,\text{mle}})_{i,j}| \le n^{-1}u_0(1+u_0^2),
\end{eqnarray}
where for a square matrix $A$, $\lambda_{\max}(A)$ denotes its largest eigenvalue and we defined
$u_0= \max\{\lambda_{\max}(\Sigma_E),  \lambda_{\max}(A_{\mathcal{E}}^\top \Omega^{-1} A_{\mathcal{E}})\}$.
\end{lemma}

\section{Applications and extensions} \label{sec:examples}

The results as presented may be extended towards slightly more general loss functions of the form $\sum_{i=1}^n \rho(\beta; Y_i, X_i)$.
When $Y_i$ has density function $f(Y_i;X_i,\beta)$, the negative log-likelihood $-\sum_{i=1}^n \log f(Y_i;X_i,\beta)$ is a natural loss function.
Other examples include of course the least squares loss function that is used for linear model estimation, but also least absolute deviation loss and the so-called Huber loss function resulting in outlier-robust estimators all fall within the scope of M-estimation.
In a regression setting one often has the form
$\ell(X\beta;Y)= \sum_{i=1}^n \rho(x_i^\top \beta; y_i)$, which we used for our main results.

We here present some other applications for selective inference for which up to now no such results are available.
This gives a flavor of the potential of this methodology.

\subsection{Modeling the location, scale and shape aspects of a distribution}

This method of selective inference also applies to the large class of flexible models, indicated by the acronym GAMLSS, that makes use of generalized additive modeling techniques for the location, scale and shape parameters of a distribution \citep{RigbyStasinopoulos2005}. Examples include
models for a  generalized Pareto distributed response with a scale and shape parameter that may both depend on covariate information. For discrete data this class of models includes the zero-inflated Poisson models for which both the mean and the parameter indicating the extra probability at zero could be modeled in terms of covariates, as well as the beta-binomial distribution which includes next to the probability of success a parameter to model possible overdispersion.
The book \citet{RigbyStasinopoulosHellerDeBastiani2019} contains a wealth of examples. See also
\citet{StasinopoulosRigbyHellerVoudourisDeBastiani2020} for more information about the models and how to work with them using the software R.


\citet{GrollHambuckersKneibUmlauf2019} proposed the use of lasso-type regularization in case of high-dimensional covariates that all enter the model in a linear way.
With the use of group lasso, however, also smooth functions could be incorporated. Let a smooth function be approximated by a spline model using $K$ basis functions, such that we can write $s_{1j}(x_{1j}) = \sum_{k=1}^K \beta_{j,k} \psi_{k}(x_{1j})$. A group lasso can treat all spline coefficients of a smooth function as one group. Also categorical covariates modeled via a number of indicator variables can be treated as a group, while continuous covariates form a group of size 1. Randomized group lasso estimation is new in this framework and proceeds via \eqref{eq:opt.problem} with inference as explained in Section~\ref{sec:inference}.

\subsection{Group lasso estimation for quasi-likelihood estimation}\label{sec:quasi}

Quasi-likelihood modeling is an attractive alternative for an explicit modeling of overdispersion since only the specification of a function for the mean response and for the variance is required.
For high-dimensional sparse models, \citet{vandeGeerMueller2012} proposed as a loss function the negative quasi-loglikelihood function, which does not need to correspond to an actual likelihood. Denoting $V$ the variance function in the model, which in case of a correct specification would yield that $\mbox{Var}(Y_i)=a_i\phi V(\mu_i)$ for a known value $a_i$, the quasi-log likelihood is defined as $\ell(\beta;Y,X) =\sum_{i=1}^n\int_{Y_i}^{\mu_i} (Y_i-u)/\{\phi a_i V(u)\}du$, see also \citet{Davison2003}. For a parametric quasi-likelihood function, $\mu_i = E[Y_i|X_i=x_i] = g(x_i^\top\beta)$ for a user-specified link function $g$, which in case of the quasi-Poisson distribution is often taken to be the logarithmic function. When adding a group lasso regularization to the loss function in addition to the added randomization, our method provides valid selective inference for the coefficients in the selected model, as explained. This novel approach to selective inference in quasi-likelihood models is implemented in Section \ref{sec:simul}.

\subsection{Quantile regression or other uses of M-estimators}

An interesting application of the use of selective inference for group lasso applications is for the situation of smoothed quantile regression \citep{HePanTanZhou2021} where we consider a setting with a large number of (grouped) variables. The introduction of a kernel density function convoluted with the non-differentiable quantile loss function leads to a globally convex loss function. Without regularization, \citet{HePanTanZhou2021} define the $\tau\in(0,1)$-smoothed quantile estimator as follows,
$
\arg\min_{\beta\in\mathbbm{R}^p} n^{-1}\sum_{i=1}^n (\rho_\tau * K_h)(Y_i-x_i^\top \beta),
$
where * denotes convolution, the check function $\rho_\tau(u) = u\{\tau-I(u<0)\}$ and $K_h(u)=K(u/h)/h$ is a scaled kernel density function with the bandwidth $h>0$.
For grouped lasso we introduce the novel regularized smoothed quantile estimator with added randomization
\begin{eqnarray*}
\widehat\beta_{n,\tau}^{(\Lambda)} \in \arg\min_{\beta\in\mathbbm{R}^p} \left\{ n^{-1/2}\sum_{i=1}^n (\rho_\tau * K_h)(Y_i-x_i^\top \beta)+
 \sum_{g\in\mathcal{G}} \lambda_g\|\beta_g\|_2  -\sqrt{n}\omega_n^\top\beta\right\},
\end{eqnarray*}
with all $\lambda_g \ge0$.
Valid selective inference for $\widehat\beta_{n,\tau}^{(\Lambda)}$, for which some of the components may be set equal to zero due to the regularization, proceeds as explained in Sections~\ref{sec:method} and \ref{sec:inference}.

\section{Simulation study} \label{sec:simul}
Section \ref{sec:simul_design} outlines the design of our simulation study of selective inference by the randomized group lasso. 
We describe the data generating process of differently distributed responses, under varying sparsity settings and how we implement our method. 
In Section \ref{sec:simul_results}, we discuss the inferential results on our simulated data.

\subsection{Design of the simulation study}\label{sec:simul_design}
Since our motivating example pertains to data with grouped (categorical) covariates, we generate a design matrix $X\in\mathbbm{R}^{n\times p}$, where $n=500$ and $p=200$, of which the columns contain both continuous and discrete grouped variables. More specifically, 120 variables are continuous and 20 are discrete with 5 levels, which are encoded into 80 binary indicators. We use the subscripts $c$ and $d$ to denote continuous and discrete variables, respectively. The continuous variables are drawn from $N_{120}(0,\Sigma)$, where $\Sigma$ follows an autoregressive of order 1 structure with $\Sigma_{i,j}=0.3^{|i-j|}$. The discrete variables are generated from a discrete uniform distribution on $[1,5]$ and one-hot encoded subsequently.

We generate four differently distributed responses: (i) Gaussian $Y_i \sim N(X_i^\top \beta,\sigma^2 I_n)$ \text{with} $\sigma=5$; (ii) Logistic $Y_i \sim Bernoulli(\frac{e^{X_i^\top \beta}}{1+e^{X_i^\top \beta}})$; (iii) Poisson $Y_i\sim Pois(e^{X_i^\top \beta})$, (iv) Negative binomial $Y_i\sim NB(r=\frac{e^{X_i^\top \beta}}{\phi-1},p=\frac{1}{\phi})$ for generating an overdispersed poisson process with expectation $e^{X_i^\top \beta}$ and overdispersion parameter $\phi=1.5$.
where $\beta \in \mathbbm{R}^{p}$ with signal size $m = \sqrt{2\tau \log(p)}$
and the number of true nonzero groups of coefficients $s=s_c+s_d$, $s \in \{5,8,10\}$ (with group sizes of 4 coefficients). In our experiment, we fix the signal size to $\tau=0.1$ ($m=1.03$) and vary $s$, the sparsity of the vector of coefficients. More specifically, we fix the number of true nonzero groups of continuous variables $s_c=3$ and vary the number of true nonzero groups of discrete variables $s_d \in \{2,5,7\}$. Every experiment is based on 500 replications. 

We compare inference using our selective asymptotic likelihood to inference from data splitting and the naive approach, as previously done in our first data example.
When data splitting, we choose from the data $(X,Y)$ a subset $(X^S,Y^S)$ of size $n_1=335$, which we use to obtain the selection $E$ by the group lasso without randomization term. 
For the case of Gaussian response, we also compare our inference method to the Bayesian approach to post-selection inference after the group lasso in \cite{PanigrahiMacDonaldKessler2021}.
As described in this earlier work, we apply a gradient based Langevin sampler to draw $2000$ samples from the proposed posterior that was formed with a diffuse prior $\pi_E = N_{|E|}(0_{|E|}, 200\  \Sigma_E)$.
After discarding the first $50$ samples as burn-in,  the appropriate quantiles of the remaining posterior samples were used to construct credible intervals.

We sample the added randomization variable $\omega_n$ in our method from $N_p(0,\Omega)$, where $\Omega=f H$ (see Section~\ref{sec:method} for the definition of $H$) and $f= \dfrac{1-r}{r}=\dfrac{n_2}{n_1} \sim 2$.
A benefit with the above choice of $\Omega$ is that we avoid inverting large matrices while solving the $|\mathcal{G}_E|$-dimensional optimization problem in \eqref{opt:esteqn}, which is our main computing step to draw selective inferences.
It is straightforward to see that
\begin{eqnarray*}
\overline{\Omega}=f\left({U}_n^{\top} H_{E,E}{U}_n\right)^{-1},
\overline{A}=\left({U}_n^{\top} H_{E,E}{U}_n\right)^{-1} {U}_n^{\top} H_{E,E}, 
\mbox{ and }
\overline{b}=-\left({U}_n^{\top} H_{E,E}{U}_n\right)^{-1} {U}_n^{\top}\Lambda_E  {u}_n.
\end{eqnarray*}
As motivated earlier, this choice of $\Omega$, in our experiments, is also made to roughly match the amount of information used for selection in the benchmark method of data splitting.
When post-selection inference is based on a likelihood model, which is the case for a Gaussian, Bernoulli, and Poisson response, we note that $K=H$ and Lemma \ref{S-lem:samplesplitting} in the Appendix suggests that we set $\Omega = f H$ where $f\sim 2$.
For the negative binomial response, we use quasi-likelihood modeling starting with a Poisson model as detailed out in Section \ref{sec:quasi} to deal with the overdispersion.
In this instance, to avoid inverting large matrices, we set $\Omega= H$ by letting $f=1$, which allows for a fair comparison of inference based on our asymptotic post-selection likelihood to inference after data splitting.

The following metrics are computed to evaluate the methods of inference. 
First, we compute the coverage rate of individual $90\%$ confidence intervals per simulation.
Second, F1-scores are computed to measure the accuracy of the model selection: 
\begin{eqnarray}
\text{F1-score}=\frac{\#\text{True Positives}}{\#\text{True Positives}+\frac{1}{2}(\#\text{False Positives}+\#\text{False Negatives})},
\label{eq:def-F1score}
\end{eqnarray}
where true/false positive/negative pertains to the selection stage (not the inference stage). 
Third, we compare the inferential power for the different methods through the averaged length of the confidence intervals.

\begin{figure}[!ht]
    \centering
    \includegraphics[width=1\linewidth]{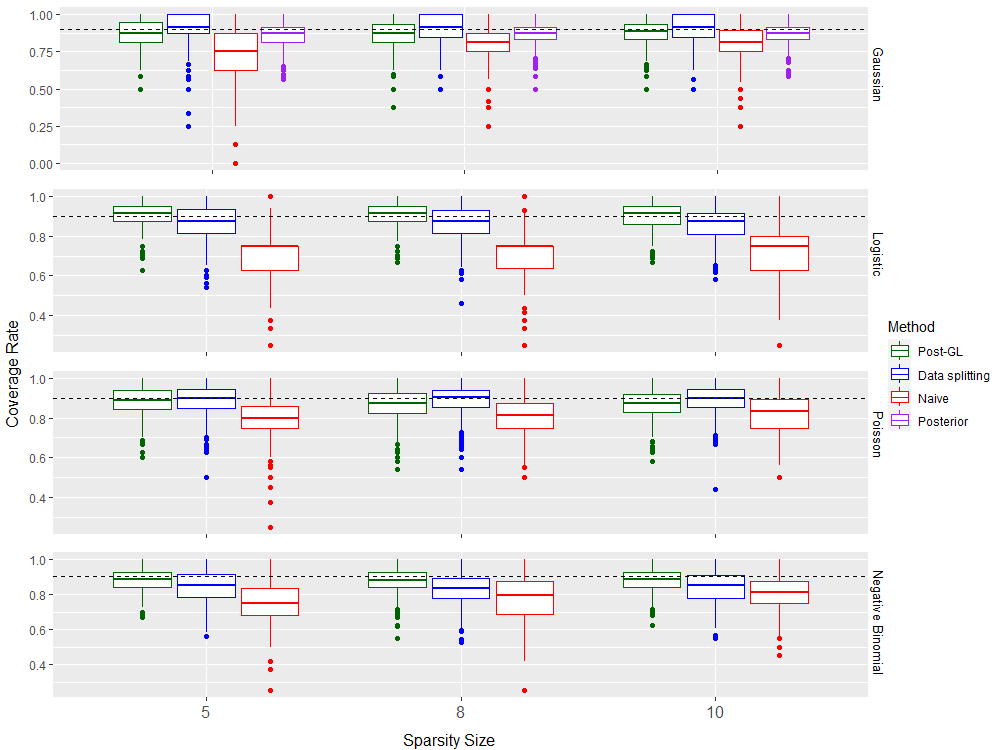}
    \caption{\small{Boxplots of coverage rate of individual $90\%$ confidence intervals for Gaussian, logistic, Poisson and negative binomial data.}}
    \label{fig:CR}
\end{figure}

\begin{figure}[!ht]
    \centering
    \includegraphics[width=1\linewidth]{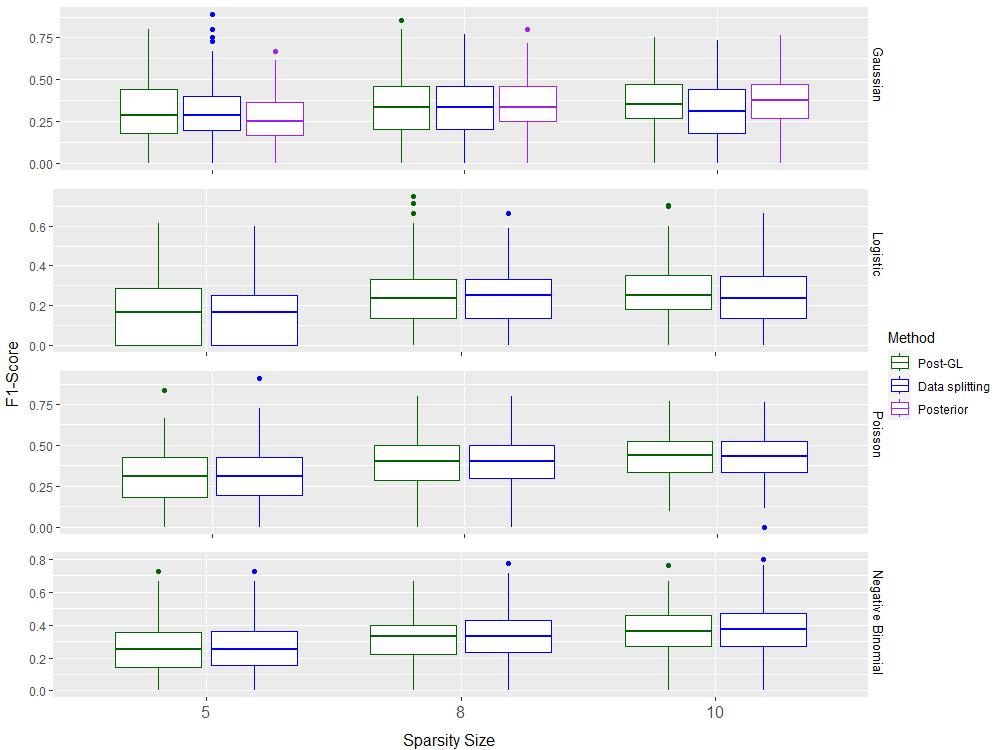}
    \caption{\small{Boxplots of F1-scores for measuring the accuracy of model selection per simulation for Gaussian, logistic, Poisson and negative binomial data.}}
    \label{fig:F1}
\end{figure}

\begin{figure}[!ht]
    \centering
    \includegraphics[width=1\linewidth]{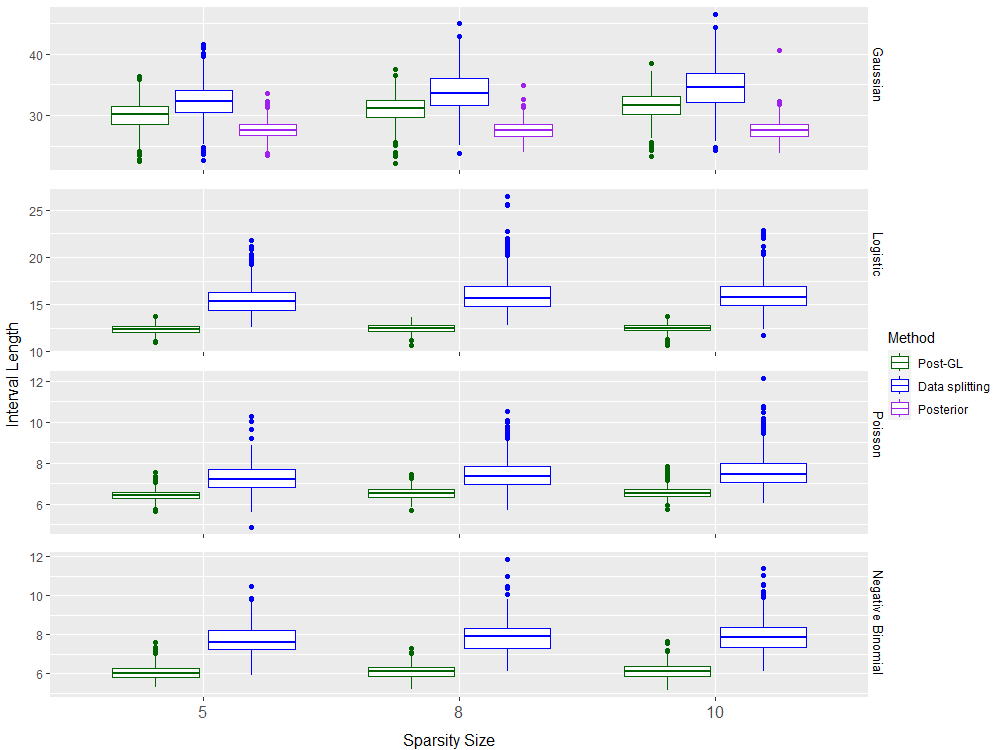}
    \caption{\small{Boxplots of average length of individual $90\%$ confidence intervals per simulation for Gaussian, logistic, Poisson and negative binomial data.}}
    \label{fig:length}
\end{figure}

\subsection{Results of the simulation study}\label{sec:simul_results}
The results of our simulation study are summarized in Figures \ref{fig:CR}, \ref{fig:F1}, and \ref{fig:length}, where the above-mentioned sparsity settings are displayed on the x-axis and inference using our asymptotic post-selection likelihood for the group lasso method is referred to using the label ``Post-GL". 
The label ``Posterior" refers to the Bayesian approach from \citet{PanigrahiMacDonaldKessler2021}, which only applies to the Gaussian case.

From Figure \ref{fig:CR}, one can conclude that confidence intervals obtained from our selective likelihood achieve the pre-specified coverage rate of $90\%$, unlike the naive confidence intervals.
Figure \ref{fig:F1}, which summarizes the F1-scores per simulation using boxplots, supports the above-mentioned statement that our choice of the randomization level allows for a fair comparison of our inference method to data splitting, since the models produced by our method and data splitting are similar in selection quality. 
Finally, Figure \ref{fig:length} shows that we obtain narrower confidence intervals than the intervals obtained after data splitting, indicating higher power of our tests. 
Our findings align with the first example in Section \ref{sec:randomization}: we obtain intervals that are valid post-selection, since they obtain the pre-specified rate of type I errors, and are simultaneously more powerful than a comparable (in selection quality) splitting method across differently generated processes and sparsity settings.

\section{Analysis of a NHANES data set} \label{sec:data}

\begin{table}[h!]\tiny
[]\label{tab:NHANES_p}
\resizebox{0.95\columnwidth}{!}{%
\begin{tabular}{|l|cc|cc|cc|}
\hline
{\color[HTML]{000000} } &
  \multicolumn{2}{c|}{{\color[HTML]{000000} Post-GL}} &
  \multicolumn{2}{c|}{{\color[HTML]{000000} Data splitting}} &
  \multicolumn{2}{c|}{{\color[HTML]{000000} Naive}} \\ \hline
 &
  \multicolumn{1}{l|}{P-value} &
  \multicolumn{1}{l|}{Group p-value} &
  \multicolumn{1}{l|}{P-value} &
  \multicolumn{1}{l|}{Group p-value} &
  \multicolumn{1}{l|}{P-value} &
  \multicolumn{1}{l|}{Group p-value}  \\ \hline
{\color[HTML]{000000} Intercept} &
  \multicolumn{1}{c|}{{\color[HTML]{FE0000} 0.600}} &
  {\color[HTML]{FE0000} 0.600} &
  \multicolumn{1}{c|}{{\color[HTML]{000000} \textbf{0.000}}} &
  {\color[HTML]{000000} \textbf{0.000}} &
  \multicolumn{1}{c|}{{\color[HTML]{FE0000} \textbf{0.000}}} &
  {\color[HTML]{FE0000} \textbf{0.000}} \\ \hline
{\color[HTML]{000000} Age} &
  \multicolumn{1}{c|}{{\color[HTML]{FE0000} 0.327}} &
  {\color[HTML]{FE0000} 0.327} &
  \multicolumn{1}{c|}{{\color[HTML]{000000} -}} &
  {\color[HTML]{000000} -} &
  \multicolumn{1}{c|}{{\color[HTML]{FE0000} \textbf{0.037}}} &
  {\color[HTML]{FE0000} \textbf{0.037}} \\ \hline
{\color[HTML]{000000} Family income to poverty ratio} &
  \multicolumn{1}{c|}{{\color[HTML]{000000} \textbf{0.000}}} &
  {\color[HTML]{000000} \textbf{0.000}} &
  \multicolumn{1}{c|}{{\color[HTML]{000000} \textbf{0.043}}} &
  {\color[HTML]{000000} \textbf{0.043}} &
  \multicolumn{1}{c|}{{\color[HTML]{000000} \textbf{0.000}}} &
  {\color[HTML]{000000} \textbf{0.000}} \\ \hline
{\color[HTML]{000000} BMI} &
  \multicolumn{1}{c|}{{\color[HTML]{FE0000} 0.616}} &
  {\color[HTML]{FE0000} 0.616} &
  \multicolumn{1}{c|}{{\color[HTML]{000000} \textbf{0.043}}} &
  {\color[HTML]{000000} \textbf{0.043}} &
  \multicolumn{1}{c|}{{\color[HTML]{FE0000} \textbf{0.093}}} &
  {\color[HTML]{FE0000} \textbf{0.093}} \\ \hline
{\color[HTML]{000000} \begin{tabular}[c]{@{}l@{}}DBD895: nr of meals not \\ prepared at home last 7 days\end{tabular}} &
  \multicolumn{1}{c|}{{\color[HTML]{000000} 0.223}} &
  {\color[HTML]{000000} 0.223} &
  \multicolumn{1}{c|}{{\color[HTML]{000000} -}} &
  {\color[HTML]{000000} -} &
  \multicolumn{1}{c|}{{\color[HTML]{000000} 0.773}} &
  {\color[HTML]{000000} 0.773} \\ \hline
{\color[HTML]{000000} \begin{tabular}[c]{@{}l@{}}DBD905: nr of ready-to-eat \\ foods in last 30 days\end{tabular}} &
  \multicolumn{1}{c|}{{\color[HTML]{000000} 0.166}} &
  {\color[HTML]{000000} 0.166} &
  \multicolumn{1}{c|}{{\color[HTML]{000000} 0.901}} &
  {\color[HTML]{000000} 0.901} &
  \multicolumn{1}{c|}{{\color[HTML]{000000} 0.171}} &
  {\color[HTML]{000000} 0.171} \\ \hline
{\color[HTML]{000000} \begin{tabular}[c]{@{}l@{}}DBD910: nr of frozen meals/\\ pizza in last 30 days\end{tabular}} &
  \multicolumn{1}{c|}{{\color[HTML]{009901} \textbf{0.000}}} &
  {\color[HTML]{009901}\textbf{0.000}} &
  \multicolumn{1}{c|}{{\color[HTML]{009901} 0.431}} &
  {\color[HTML]{009901} 0.431} &
  \multicolumn{1}{c|}{{\color[HTML]{000000} \textbf{0.000}}} &
  {\color[HTML]{000000} \textbf{0.000}} \\ \hline
{\color[HTML]{000000} Race (ref.: Other)} &
  \multicolumn{1}{c|}{} &
   &
  \multicolumn{1}{c|}{} &
   &
  \multicolumn{1}{c|}{} &
   \\ \hline
\multicolumn{1}{|r|}{Mexican american} &
  \multicolumn{1}{c|}{{\color[HTML]{009901} \textbf{0.032}}} &
  {\color[HTML]{009901} } &
  \multicolumn{1}{c|}{{\color[HTML]{009901} -}} &
  {\color[HTML]{009901} } &
  \multicolumn{1}{c|}{0.262} &
   \\ \cline{1-2} \cline{4-4} \cline{6-6}
\multicolumn{1}{|r|}{Other hispanic} &
  \multicolumn{1}{c|}{0.869} &
  {\color[HTML]{009901} } &
  \multicolumn{1}{c|}{-} &
  {\color[HTML]{009901} } &
  \multicolumn{1}{c|}{0.814} &
   \\ \cline{1-2} \cline{4-4} \cline{6-6}
\multicolumn{1}{|r|}{Non-hispanic white} &
  \multicolumn{1}{c|}{0.325} &
  {\color[HTML]{009901} } &
  \multicolumn{1}{c|}{-} &
  {\color[HTML]{009901} } &
  \multicolumn{1}{c|}{0.953} &
   \\ \cline{1-2} \cline{4-4} \cline{6-6}
\multicolumn{1}{|r|}{Non-hispanic black} &
  \multicolumn{1}{c|}{{\color[HTML]{FE0000} 0.159}} &
  \multirow{-4}{*}{{\color[HTML]{009901} \textbf{0.000}}} &
  \multicolumn{1}{c|}{-} &
  \multirow{-4}{*}{{\color[HTML]{009901} -}} &
  \multicolumn{1}{c|}{{\color[HTML]{FE0000} \textbf{0.002}}} &
  \multirow{-4}{*}{\textbf{0.000}} \\ \hline
{\color[HTML]{000000} Civil state (ref.: Never married)} &
  \multicolumn{1}{c|}{{\color[HTML]{000000} }} &
  {\color[HTML]{000000} } &
  \multicolumn{1}{c|}{{\color[HTML]{000000} }} &
  {\color[HTML]{000000} } &
  \multicolumn{1}{c|}{{\color[HTML]{000000} }} &
  {\color[HTML]{000000} } \\ \hline
\multicolumn{1}{|r|}{{\color[HTML]{000000} Married/Living with}} &
  \multicolumn{1}{c|}{{\color[HTML]{FE0000} 0.133}} &
  {\color[HTML]{000000} } &
  \multicolumn{1}{c|}{{\color[HTML]{000000} 0.171}} &
  {\color[HTML]{000000} } &
  \multicolumn{1}{c|}{{\color[HTML]{FE0000} \textbf{0.016}}} &
  {\color[HTML]{000000} } \\ \cline{1-2} \cline{4-4} \cline{6-6}
\multicolumn{1}{|r|}{{\color[HTML]{000000} Widowed/Divorced/Separated}} &
  \multicolumn{1}{c|}{{\color[HTML]{000000} 0.315}} &
  \multirow{-2}{*}{{\color[HTML]{000000} \textbf{0.000}}} &
  \multicolumn{1}{c|}{{\color[HTML]{000000} 0.202}} &
  \multirow{-2}{*}{{\color[HTML]{000000} \textbf{0.000}}} &
  \multicolumn{1}{c|}{{\color[HTML]{000000} 0.639}} &
  \multirow{-2}{*}{{\color[HTML]{000000} \textbf{0.000}}} \\ \hline
{\color[HTML]{000000} Healthy diet (ref.: Poor)} &
  \multicolumn{1}{c|}{{\color[HTML]{000000} }} &
  \multicolumn{1}{l|}{{\color[HTML]{000000} }} &
  \multicolumn{1}{c|}{{\color[HTML]{000000} }} &
  \multicolumn{1}{l|}{{\color[HTML]{000000} }} &
  \multicolumn{1}{c|}{{\color[HTML]{000000} }} &
  \multicolumn{1}{l|}{{\color[HTML]{000000} }} \\ \hline
\multicolumn{1}{|r|}{{\color[HTML]{000000} Excellent}} &
  \multicolumn{1}{c|}{{\color[HTML]{009901} \textbf{0.023}}} &
  {\color[HTML]{000000} } &
  \multicolumn{1}{c|}{{\color[HTML]{009901} 0.361}} &
  {\color[HTML]{000000} } &
  \multicolumn{1}{c|}{{\color[HTML]{000000} \textbf{0.000}}} &
  {\color[HTML]{000000} } \\ \cline{1-2} \cline{4-4} \cline{6-6}
\multicolumn{1}{|r|}{{\color[HTML]{000000} Very good}} &
  \multicolumn{1}{c|}{{\color[HTML]{000000} \textbf{0.002}}} &
  {\color[HTML]{000000} } &
  \multicolumn{1}{c|}{{\color[HTML]{000000} \textbf{0.002}}} &
  {\color[HTML]{000000} } &
  \multicolumn{1}{c|}{{\color[HTML]{000000} \textbf{0.000}}} &
  {\color[HTML]{000000} } \\ \cline{1-2} \cline{4-4} \cline{6-6}
\multicolumn{1}{|r|}{{\color[HTML]{000000} Good}} &
  \multicolumn{1}{c|}{{\color[HTML]{000000} \textbf{0.001}}} &
  {\color[HTML]{000000} } &
  \multicolumn{1}{c|}{{\color[HTML]{000000} \textbf{0.001}}} &
  {\color[HTML]{000000} } &
  \multicolumn{1}{c|}{{\color[HTML]{000000} \textbf{0.000}}} &
  {\color[HTML]{000000} } \\ \cline{1-2} \cline{4-4} \cline{6-6}
\multicolumn{1}{|r|}{{\color[HTML]{000000} Fair}} &
  \multicolumn{1}{c|}{{\color[HTML]{009901} \textbf{0.005}}} &
  \multirow{-4}{*}{{\color[HTML]{000000} \textbf{0.000}}} &
  \multicolumn{1}{c|}{{\color[HTML]{009901} 0.123}} &
  \multirow{-4}{*}{{\color[HTML]{000000} \textbf{0.000}}} &
  \multicolumn{1}{c|}{{\color[HTML]{000000} \textbf{0.000}}} &
  \multirow{-4}{*}{{\color[HTML]{000000} \textbf{0.000}}} \\ \hline
{\color[HTML]{000000} Male} &
  \multicolumn{1}{c|}{{\color[HTML]{FE0000} 0.822}} &
  {\color[HTML]{FE0000} 0.822} &
  \multicolumn{1}{c|}{{\color[HTML]{000000} \textbf{0.028}}} &
  {\color[HTML]{000000} \textbf{0.028}} &
  \multicolumn{1}{c|}{{\color[HTML]{FE0000} \textbf{0.000}}} &
  {\color[HTML]{FE0000} \textbf{0.000}} \\ \hline
{\color[HTML]{000000} Daily 4/5 or more drinks} &
  \multicolumn{1}{c|}{{\color[HTML]{FE0000} 0.141}} &
  {\color[HTML]{FE0000} 0.141} &
  \multicolumn{1}{c|}{{\color[HTML]{000000} \textbf{0.028}}} &
  {\color[HTML]{000000} \textbf{0.028}} &
  \multicolumn{1}{c|}{{\color[HTML]{FE0000} \textbf{0.000}}} &
  {\color[HTML]{FE0000} \textbf{0.000}} \\ \hline
{\color[HTML]{000000} High blood pressure/Hypertension} &
  \multicolumn{1}{c|}{{\color[HTML]{FE0000} 0.449}} &
  {\color[HTML]{FE0000} 0.449} &
  \multicolumn{1}{c|}{{\color[HTML]{000000} \textbf{0.026}}} &
  {\color[HTML]{000000} \textbf{0.026}} &
  \multicolumn{1}{c|}{{\color[HTML]{FE0000} \textbf{0.000}}} &
  {\color[HTML]{FE0000} \textbf{0.000}} \\ \hline
{\color[HTML]{000000} Smoked at least 100 cigarettes in life} &
  \multicolumn{1}{c|}{{\color[HTML]{FE0000} 0.180}} &
  {\color[HTML]{FE0000} 0.180} &
  \multicolumn{1}{c|}{{\color[HTML]{000000} 0.617}} &
  {\color[HTML]{000000} 0.617} &
  \multicolumn{1}{c|}{{\color[HTML]{FE0000} \textbf{0.018}}} &
  {\color[HTML]{FE0000} \textbf{0.018}} \\ \hline
{\color[HTML]{000000} Asthma} &
  \multicolumn{1}{c|}{{\color[HTML]{009901} \textbf{0.000}}} &
  {\color[HTML]{009901} \textbf{0.000}} &
  \multicolumn{1}{c|}{{\color[HTML]{009901} 0.514}} &
  {\color[HTML]{009901} 0.514} &
  \multicolumn{1}{c|}{{\color[HTML]{000000} \textbf{0.000}}} &
  {\color[HTML]{000000} \textbf{0.000}} \\ \hline
\end{tabular}%
}\caption{\small{P-values for effects on depression from our post-selection asymptotic likelihood, data splitting and naive inference. On a $90\%$ level, cases where significant effects were found by naive inference, but not selective inference are in red. Cases where significant effects were found by selective inference, but not data splitting are indicated in green.}}
\end{table}

In this section, our method is applied to a dataset compiled from the National Health and Nutrition Examination Survey  (NHANES) round of 2017-March 2020 pre-pandemic \citep{NHANES2017_2020}. 
The NHANES dataset contains demographic, nutritional and health related data about United States residents. We combined data from both its questionnaires and physical examinations to get to a subset of size $n\times p=5985\times 50$. The selection of this subset of variables from the NHANES was done in a data independent way motivated by literature: we compiled a similar dataset as \cite{LiLingli2022Abel} for investigating the incidence of depression. Similarly as \cite{LiLingli2022Abel}, we create a binary indicator of depression where scores from zero to three on nine depression screening items were summed and dichotomized into not depressed (scores 0-9) and depressed (scores 10-27).
There are $24$ covariates in the data, with $8$ of them being grouped categorical covariates that have been one-hot encoded (race, education, marital status, income, alcohol consumption, diet health, milk consumption, and diabetes), while the rest are continuous covariates. 
We use group lasso, which can handle this common type of dataset, to select (grouped) variables for estimating the incidence of depression. A non-penalized generalized linear model with logistic link function $h$,
$
     \ell (X\beta;Y) = - \sum_{i=1}^n [y_i \log h(x_i^{\top} \beta) + (1-y_i) \log \{1-h(x_i^{\top} \beta)\}] 
$ is refitted after the selection to estimate the probability of belonging to the depression group given the selected covariates. Similarly to the simulation study, we compare three methods of inference to assess the statistical significance of the regression coefficients. Naive inference selects variables without randomization and uses the same data for statistical inference. In data splitting, a randomly selected subset of $90\%$ of the data is used for selection (without randomization), and statistical inference is obtained based on the remaining data. Finally, we conduct inference based on our asymptotic post-selection likelihood after randomized selection by solving \eqref{eq:opt.problem} with $\omega_n \sim N(0,\frac{1}{9}\widehat{H})$. This analysis is conducted for a regularization parameter $\lambda=0.5$ where the regularization parameter for the group of covariates $g$ is $\lambda_{g,n}=\lambda
\{ |g|/(\sum_{g\in\mc{G}}|g|/|\mc{G}|) n \widehat{\text Var}(Y) 2 \log(p)\}^{1/2}$.

\begin{figure}[!ht]
    \centering
    \includegraphics[width=0.8\textwidth]{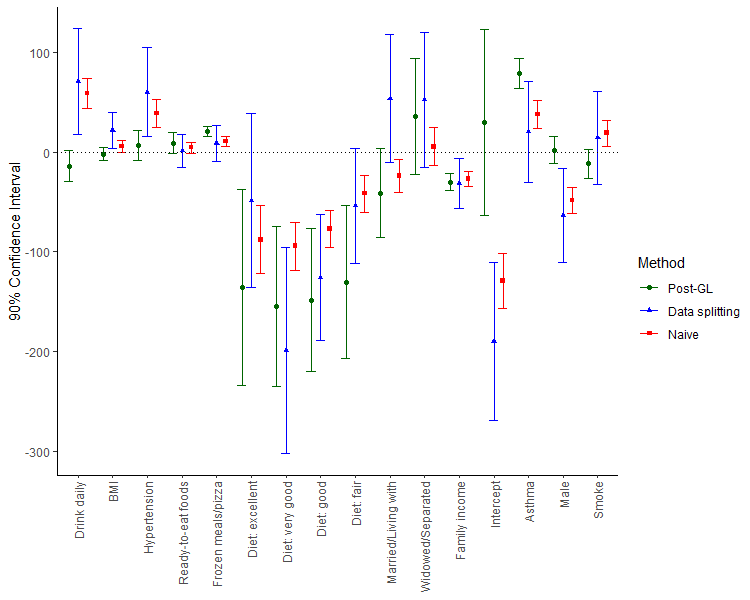}
    \caption{\small{$90\%$ confidence intervals post-selection by the group lasso for estimating the incidence of depression: from our post-selection likelihood, data splitting and naive inference.}}
    \label{fig:NHANES_CI_plot}
\end{figure}

Table \ref{tab:NHANES_p} gives the p-values for individual coefficients and at the group level and Figure \ref{fig:NHANES_CI_plot}
plots the corresponding $90\%$ confidence intervals for individual coefficients (that were selected by all methods). For confidence intervals of all selected regression coefficients we refer to Table \ref{S-tab:NHANES_CI} in the Appendix. The intervals obtained from data splitting are wider, indicating lower power, than the intervals obtained from our asymptotic post-selection likelihood. For example, we detected a significant negative association between ``excellent" diet health (compared to poor) and the probability of depression, which was not detected by inference after data splitting. On the other hand, naive inference finds in nine cases significant effects which are not significant when controlling for the selection (see the effect of the intercept, age, BMI, non-hispanic black, married or living with a partner, gender, alcohol consumption, hypertension and smoking).

\section{Discussion} \label{sec:disc}

The main contribution of this work on selective inference is that we step away from models for Gaussian distributions with least squares loss. Likelihood-based models, including generalized linear and generalized additive models can be dealt with, as are estimation methods using quasi-likelihood estimating equations. Second, the regularized methods that we study may include $l_2$-norms for grouped covariates, implying that we cannot rely on polyhedral results for selective inference. Third, by including an extra randomization during the regularized estimation, the selective inference procedure results in confidence regions for which we show that they possess finite volumes, in contrast to unrandomized selective inference where the expected lengths of confidence intervals can be infinite.

While in this paper we worked under the assumption of a large but fixed number of covariates $p$, for future research it is of interest to extend these results towards a growing number of covariates.
\cite{SurCandes2019, ZhouClaeskens2022} have shown that the classical theory of maximum likelihood estimation is not valid anymore in high-dimensional logistic regression models. 
Our focus diverges from this line of work as we do not infer for a $p$-dimensional parameter vector. 
Instead, we infer for a post-selection parameter vector with a relatively low dimension, which is defined in Equation \eqref{eqn:targetM}. 
Nonetheless, for these asymptotic scenarios, we believe that a novel theoretical approach is necessary to extend the current theory. 

One of the significant challenges is obtaining a correction for the selection effect in the growing $p$ regime.
In our current approach, the solution demands a large deviation-type result used in Theorem \ref{Th:LDP} to achieve a limiting approximation to the probability of the selection event. 
In the field of probability theory, recent progress has been made towards obtaining large deviation results for growing dimensions by projecting them to lower-dimensional subspaces. 
See, for example, papers by \cite{kim2022asymptotic, kim2023large}.
We leave exploring the potential of these new developments for providing selective inferences to future work.

\section*{Acknowledgements}
S.Panigrahi acknowledges the support of NSF grants 1951980 and 2113342.
G.Claeskens acknowledges the support of the Research Foundation Flanders and KU Leuven Research Fund C1-project C16/20/002.

\biblist

\appendix
\section{Appendix- Proofs of theorems}

\subsection{Auxiliary Results}
\begin{lemma}[Jacobian of Change-of-Variables]
	\label{lem:Simplified Jacobian of CoV}
	Suppose $\Pi^*$ is the mapping in \eqref{Pi:star}.
	Then,
	\begin{align*}
 	\det D_{\Pi^*}(\sqrt{n}\beta_{n,E}, \sqrt{n}\beta_{n,E}^\perp,\sqrt{n}{\gamma}, u_n, z_n)
 	=
 	\det D_{\Pi}(\sqrt{n}{\gamma}, u_n, z_n).
 \end{align*}\end{lemma}
 \begin{proof}
 	From the definition of $\Pi^*$, note that
 	\begin{align*}
 		D_{\Pi^*}(b, b^\perp, g, u, z) =
 		\begin{pmatrix}
 			I_{|E|} &
 			0 &
 			0\\
 			0 &
 			I_{|E'|} &
 			0\\
 			\frac{\partial \Pi^*_3}{\partial b} &
 			\frac{\partial \Pi^*_3}{\partial b^\perp} &
 			\frac{\partial \Pi^*_3}{\partial (g,u,z)}
 		\end{pmatrix},
 	\end{align*}
 where
 $\Pi_3^*( b, b^\perp, \omega) = \Pi_{b, b^\perp}^{-1} (\omega)=(g^\top,u^\top,z^\top)^\top.$
Therefore,
 	\begin{align*}
 		\det D_{\Pi^*}(b, b^\perp, g, u, z) = \det\begin{pmatrix}
 			\frac{\partial \Pi^*_3}{\partial (g,u,z)}
 		\end{pmatrix}
 		= \det D_{\Pi}(g,u,z).
 	\end{align*}
 \end{proof}

\begin{lemma}[Calculating the Jacobian Term]
\label{lem:JacobianComputation}
Let $D_{\Pi}$ be as defined in Proposition 2.2
and define $b_{E,n}=|\det(\Lambda_{E'})/\det(H_{E,E})|$.
Then, we have
\begin{align*}
		|\det D_{\Pi}(\sqrt{n}\gamma_n, u_n, z_n)|
		= J_{\Pi}(\sqrt{n}\gamma_n, u_n) \defeq  b_{E,n} \operatorname{det}\left(\Gamma_n+\bar{U}_n^{\top} H_{E, E}^{-1} \Lambda_E \bar{U}_n\right).
\end{align*}

\begin{proof}
First, to derive $D_{\Pi}(\sqrt{n}{\gamma}, u_n, z_n)$, note that
the $p$-vector
$$
\Pi_{b, b^{\perp}}(g,u,z)
=  A_{\mathcal{E}} b + B_{\mathcal{E}} g+ C_{\mathcal{E}} (\Lambda_{E'}z + b^{\perp}) + D_{\mathcal{E}}u,
$$
can be split into the first $|E|$ rows, denoted $\Pi^{(1)}_{b,b^\perp}(g,u,z)$ and the last $|E'|$ rows, denoted $\Pi^{(2)}_{b,b^\perp}(g,u,z)$. Using the definition of $\Sigma_E$, this results in
\begin{eqnarray*}
\Pi^{(1)}_{b,b^\perp}(g,u,z) & = & - K_{E,E}H_{E,E}^{-1}\Sigma_{E}^{-1} b + H_{E,E} \widehat{U}_n g + 0_{|E|} + \Lambda_E u \\
\Pi^{(2)}_{b,b^\perp}(g,u,z) & = & - K_{E',E}H_{E,E}^{-1}\Sigma_{E}^{-1}b + H_{E',E}\widehat{U}_n g +(\Lambda_{E'}z+b^\perp) +  0_{|E'|}.
\end{eqnarray*}
Consequently, the derivative matrix
\begin{align*}
	D_{\Pi}(\sqrt{n}\gamma_n, {u}_n,  {z}_n)=\left(\begin{array}{ccc}
\frac{\partial \Pi^{(1)}}{\partial \sqrt{n}\gamma_n} & \frac{\partial \Pi^{(1)}}{\partial u_n} & \frac{\partial \Pi^{(1)}}{\partial z_n} \\
\frac{\partial \Pi^{(2)}}{\partial \sqrt{n}\gamma_n} & \frac{\partial \Pi^{(2)}}{\partial u_n} & \frac{\partial \Pi^{(2)}}{\partial z_n}
\end{array}\right).
\end{align*}
To evaluate the determinant of this matrix,
one may consider the following composition of functions $\psi, \xi$ such
that $\Pi_{b,b^\perp}(g,u,z)=\xi\circ \psi(g,u,z)$.
Hereby is
\begin{eqnarray*}
\psi & = & (\psi_1,\psi_2): \mb{R}^{|\mc{G}_E|}\times \mb{R}^{|E|}\times \mb{R}^{|E'|}\to \mb{R}^{|E|}\times\mb{R}^{|E'|}: (g,u,z)\mapsto (\text{bd}(u)g,z) \\
\xi & = & (\xi_1,\xi_2): \mb{R}^{|E|}\times \mb{R}^{|E'|}\to \mb{R}^{|E|}\times \mb{R}^{|E'|}: (b^{(\Lambda)},z)\mapsto \big(\xi_1(b^{(\Lambda)},z), \xi_2(b^{(\Lambda)},z)\big)
\end{eqnarray*}
with function values
\begin{eqnarray*}
\xi_1(b^{(\Lambda)},z) & = &  H_{E,E}b^{(\Lambda)}-K_{E,E}H_{E,E}^{-1}\Sigma_E^{-1} b+\Lambda_E u \\
\xi_2(b^{(\Lambda)},z) & = &  H_{E',E}b^{(\Lambda)}-K_{E',E}H_{E,E}^{-1}\Sigma_E^{-1} b+b^\perp+\Lambda_{E'} z.
\end{eqnarray*}
Consequently,
$ 
	|\det D_{\Pi}|=|\det D_{\xi}|\cdot|\det D_{\psi}|.
$ 
First, to calculate $\det D_{\xi}$, note
\begin{align*}
	\det D_{\xi}(\sqrt{n}\widehat\beta_{n,E}^{(\Lambda)}, z_n)
	=
	\det
	\begin{pmatrix}
		\dd{\xi_1}{\sqrt{n}\widehat\beta_{n,E}^{(\Lambda)}}
		&
		\dd{\xi_1}{z_n}\\
		\dd{\xi_2}{\sqrt{n}\widehat\beta_{n,E}^{(\Lambda)}}
		&
		\dd{\xi_2}{z_n}
	\end{pmatrix}
	=
	\det \big( \dd{\xi_1}{\sqrt{n}\widehat\beta_{n,E}^{(\Lambda)}} \big)
	\cdot \det(\Lambda_{E'}).
\end{align*}
Hence, it suffice to calculate $\dd{\xi_1}{\sqrt{n}\widehat\beta_{n,E}^{(\Lambda)}}$, where we explicitly write the dependence of $\widehat{u}_n$ on the group lasso estimator $\widehat\beta_{n,E}^{(\Lambda)}$,
\begin{align*}
	\dd{\xi_1}{\sqrt{n}\widehat\beta_{n,E}^{(\Lambda)}} =& H_{E,E}
		+
\dd{}{\sqrt{n}\widehat\beta_{n,E}^{(\Lambda)}}\left(\left(\lambda_g u_{n,g}(\sqrt{n}\widehat\beta_{n,E}^{(\Lambda)})\right)_{g \in \mathcal{G}_E}\right)\\
		=& H_{E,E} + \Lambda_E\bar{\Gamma}_n^{-1} \left(I_{|E|} - \widehat{U}_n\widehat{U}_n^\top\right), \text{ where } \bar\Gamma_n=\operatorname{diag}\left(\left(\sqrt{n}\gamma_{n,g} 1_{|g|}\right)_{g \in \mathcal{G}_E}\right).
\end{align*}

For the other determinant $\det D_{\psi}$, we notice
\begin{align*}
	\det D_\psi(\sqrt{n} \gamma_n, {u}_n,  {z}_n)
	=
	\det
	\left(\begin{array}{ccc}
	\frac{\partial \psi_1}{\partial \sqrt{n}\gamma_n} &
	\frac{\partial \psi_1}{\partial u_n} &
	\frac{\partial \psi_1}{\partial z_n} \\
	\frac{\partial \psi_2}{\partial \sqrt{n}\gamma_n} &
	\frac{\partial \psi_2}{\partial u_n} &
	\frac{\partial \psi_2}{\partial z_n}
\end{array}\right)
	=
	\det
	\begin{pmatrix}
		\frac{\partial \psi_1}{\partial \sqrt{n}\gamma_n}&
		\frac{\partial \psi_1}{\partial u_n}
	\end{pmatrix},
\end{align*}
and it follows that $\frac{\partial \psi_1}{\partial \sqrt{n}\gamma_n} = \widehat U_n$. To evaluate $\frac{\partial \psi_1}{\partial u_n}$, consider the tangent space $$
T_{u_{n,g}} \mathcal{S}^{|g|-1}=\left\{v\in \mb{R}^{|g|}: v^{\top} u_{n,g}=0\right\}
$$
at $u_{n,g} \in \mathcal{S}^{|g|-1}$, the $(|g|-1)$-dimensional unit sphere which is embedded in $\mb{R}^{|g|}$.
We denote by $\bar{U}_{n,g}$ the matrix of dimension
$|g|\times(|g|-1)$ of which the columns form an orthonormal basis for this tangent space, hence each column is orthogonal with $u_{n,g}$. These matrices are combined in a matrix $\bar U_n$ with dimension $|E|\times (|E|-|\mc{G}_E|)$, which is defined as
$\bar U_n = \text{block diag}(U_{n,g}; g\in \mc{G}_E)$.

 To differentiate any differentiable $h$ at $u_{n,g}$, we have  \citep[e.g.,][Ch.~3]{Lee2003}
\begin{align*}
\frac{\partial h\left(u_{n,g}\right)}{\partial u_{n,g}}=\frac{\partial h\left(u_{n,g}+\bar{U}_{n,g} y_g\right)}{\partial y_g},
\end{align*}
for $y_g \in \mb{R}^{|g|-1}$ and therefore, with $\psi_1(\sqrt{n}\gamma_n,u_n,z_m)=\widehat U_n\sqrt{n}\gamma_n = \sqrt{n}\widehat{\beta}_{n,E}^{(\Lambda)}$,
it holds that $\partial \psi_1/(\partial u_{n}) = \bar\Gamma_n \bar U_n$.
Using this expression, we deduce
\begin{align*}
	\det D_{\psi} &= \det \begin{pmatrix}
		\frac{\partial \psi_1}{\partial \sqrt{n}\gamma_n}&
		\frac{\partial \psi_1}{\partial u_n}
	\end{pmatrix}
	=
	\det\begin{pmatrix}
		\widehat{U}_n & \bar\Gamma_n \bar U_n
	\end{pmatrix} = (-1)^{|\mc{G}_E|(|E|-|\mc{G}_E|)}
	\det\begin{pmatrix}
		\bar\Gamma_n \bar U_n & \widehat{U}_n
	\end{pmatrix} \\
	& = a_{E,n}
	\det
	\begin{pmatrix}
		\begin{pmatrix}
		\bar U_n^\top \\
		\widehat{U}_n^\top
		\end{pmatrix}
		\begin{pmatrix}
		\bar\Gamma_n \bar U_n & \widehat U_n
		\end{pmatrix}
	\end{pmatrix}
 = a_{E,n}
	\det
	\begin{pmatrix}
		\Gamma_n & 0\\
		0 & I_{|\mc{G}_E|}
	\end{pmatrix}
	= a_{E,n}\det\Gamma_n,
\end{align*}
with $a_{E,n} =  (-1)^{|\mc{G}_E|(|E|-|\mc{G}_E|)}/\det(\bar U_n \; \widehat{U}_n)$ and
where we used the identity $\bar U_n^\top \bar\Gamma_n \bar U_n = \Gamma_n$ in the second line.
Since $(\bar U_n \; \widehat{U}_n)^\top = (\bar U_n \; \widehat{U}_n)^{-1}$, it holds that
$\det(\bar U_n \; \widehat{U}_n)=\pm 1$.
Therefore, with $b_{E,n}=|\det(\Lambda_{E'})/\det(H_{E,E})|$,
and
\begin{align*}
	\big|\det D_{\Pi}\big|
=&\big|\det D_{\lambda}\big| \cdot \big|\det D_{\psi}\big|\\
	= & |a_{E,n}\det{\Lambda_{E'}}|\big|\det\left(H_{E,E} + \Lambda_{E}\bar{\Gamma}_n^{-1} \left(I_{|E|} - U_nU_n^\top\right)\right)\det \Gamma_n \big|\\
	= & b_{E,n}
 \bigg|\det\left(I_{|E|} + H_{E,E}^{-1}\Lambda_{E}\bar{\Gamma}_n^{-1} \left(I - U_nU_n^\top\right)\right)\det \Gamma_n \bigg| \\
   = & b_{E,n}\bigg|\det\left(\bar{U}_n^\top \bar{U}_n + \bar{U}_n^\top H_{E,E}^{-1}\Lambda_{E}\bar{\Gamma}_n^{-1} \left(\bar{U}_n - U_nU_n^\top\bar{U}_n\right)\right)\det \Gamma_n\bigg|\\
	= & b_{E,n}  \bigg|\det\left(I_{|E|-|\mc{G}_E|} + \bar{U}_n^\top H_{E,E}^{-1}\Lambda_{E}\bar{\Gamma}_n^{-1}  \bar{U}_n  \right)\det \Gamma_n\bigg|\\
	= & b_{E,n}\bigg|\det\left(\Gamma_n + \bar{U}_n^\top H_{E,E}^{-1}\Lambda_{E}  \bar{U}_n \right) \bigg|.
\end{align*}
Thus, we obtain
$ 
	\big|\det D_{\Pi}\big| = b_{E,n} \det(\Gamma_n + \bar{U}_n^\top H_{E,E}^{-1}\Lambda_E\bar{U}_n) = b_{E,n} J_{\Pi}(\sqrt{n}\gamma_n, u_n).
$ 
\end{proof}
\end{lemma}

\begin{lemma}
Define the partition $\{M_g\}_{g\in \mc{G}_E}$of indices $\{1,2,\ldots,|E|-|\mc{G}_E|\}$	 such that $M_g$ is the set of $|g|-1$ indices along the diagonal of $\Gamma_n$ that corresponds to group $g$.
Then
\begin{align*}
	\dd{J_{\Pi}(\sqrt{n}\gamma_n, u_n)}{\sqrt{n}\gamma_{n,g}} =  \sum_{i\in M_g} [(\Gamma_n+\bar{U}_n^{\top} H_{E,E} ^{-1} \Lambda_E \bar{U}_n)^{-1}]_{ii}.
\end{align*}
\end{lemma}
\begin{proof}
	First we note the following identity \citep[Ch.~15, eqn (8.8)]{Harville1997}: for a $r\times r$ matrix $X$, with $i, j\in \{1,\ldots,r\}$,
		$ {\partial \log \operatorname{det}(X)}/{\partial X_{ij}} =(X^{-1})_{ji}$.
	Then we consider differentiating the following composite function
	\begin{align*}
		\sqrt{n}\gamma_{n,g} \stackrel{J_1}{\longmapsto}\left(\Gamma_n+\bar{U}_n^{\top}H_{E,E}^{-1} \Lambda_E \bar{U}_n\right) \stackrel{J_2}{\longmapsto} \log \operatorname{det}\left(\Gamma_n+\bar{U}_n^{\top}H_{E,E}^{-1} \Lambda_E \bar{U}_n\right).
	\end{align*}
	Then, notice
	\begin{align*}
		\dd{J_2\circ J_1}{\sqrt{n}\gamma_{n,g}}
		=& \sum_{i,j}
		\dd{\log \operatorname{det}\left(\Gamma_n+\bar{U}_n^{\top}H_{E,E}^{-1} \Lambda_E \bar{U}_n\right)}{\left(\Gamma_n+\bar{U}_n^{\top}H_{E,E}^{-1} \Lambda_E \bar{U}_n\right)_{ij}}
		\dd{\left(\Gamma_n+\bar{U}^{\top}H_{E,E}^{-1} \Lambda_E \bar{U}_n\right)_{ij}}{\sqrt{n}\gamma_{n,g}}\\
		=& \sum_{i,j}[(\Gamma_n+\bar{U}_n^{\top} H_{E,E} ^{-1} \Lambda_E  \bar{U}_n)^{-1}]_{ji}
		\dd{\left(\Gamma_n+\bar{U}_n^{\top}H_{E,E}^{-1} \Lambda_E \bar{U}_n\right)_{ij}}{\sqrt{n}\gamma_{n,g}}
	\end{align*}
	by the chain rule, while
	\begin{align*}
		\dd{\left(\Gamma_n+\bar{U}_n^{\top}H_{E,E}^{-1} \Lambda_E \bar{U}_n\right)_{ij}}{\sqrt{n}\gamma_{n,g}} =
		\begin{cases}
			1, 	\ i=j, i \in M_g,\\
			0, 	\ \text{otherwise.}
		\end{cases}
	\end{align*}
	Therefore,
	\begin{align*}
		\dd{J_{\Pi}(\sqrt{n}\gamma_n, u_n)}{\sqrt{n}\gamma_{n,g}}
		=  \dd{J_2\circ J_1}{\sqrt{n}\gamma_{n,g}}
		=  \sum_{i\in M_g} [(\Gamma_n+\bar{U}_n^{\top} H_{E,E} ^{-1} \Lambda_E \bar{U}_n)^{-1}]_{ii}.
	\end{align*}
\end{proof}

\begin{lemma}
The $(g,g')-$th entry, with $g,g'\in \{1,\ldots,|\mc{G}_E|\}$, of the Hessian, the second derivative with respect to the first component vector evaluated at $\sqrt{n}\gamma_n$, of $\log J_{\Pi}(\sqrt{n}\gamma_n, u_n)$ satisfies the following relation
\begin{align*}
	\left( \nabla^2 \log J_{\Pi}(\sqrt{n}\gamma_n, u_n)\right)_{g,g'}
	= - \sum_{i\in M_g}\sum_{j\in M_{g'}} [(\Gamma_n+\bar{U}_n^{\top} H_{E,E} ^{-1} \Lambda_E \bar{U}_n)^{-1}]_{ij}[(\Gamma_n+\bar{U}_n^{\top} H_{E,E} ^{-1} \Lambda_E \bar{U}_n)^{-1}]_{ji}.
\end{align*}
\end{lemma}
\begin{proof}
From \citet[][Ch.~15]{Harville1997}, for a $r\times r$ matrix $X$, and $i,j \in \{1,\ldots,r\}$,
$ {\partial X^{-1}}/{\partial X_{ij}} = - X^{-1} e_i e_j^\top X^{-1}$
where $e_i$ is a unit vector with a 1 at the $i$th place and zeros elsewhere.
Hence for its $k,l$th element it holds that
$
		{\partial\left(X^{-1}\right)_{k l}}/{\partial X_{i j}}=-\left(X^{-1}\right)_{k i}\left(X^{-1}\right)_{j l}.
$
	Now note that
	\begin{align*}
& \left( \nabla^2 \log J_{\Pi}(\sqrt{n}\gamma_n, u_n)\right)_{g,g'}
		=
		\dd{}{\sqrt{n}\gamma_{n,g'}}\dd{J_{\Pi}(\sqrt{n}\gamma_n, u_n)}{\sqrt{n}\gamma_{n,g}} \\
		=& \dd{}{\sqrt{n}\gamma_{n,g'}} \sum_{i\in M_g} [(\Gamma_n+\bar{U}_n^{\top} H_{E,E} ^{-1} \Lambda_{E} \bar{U}_n)^{-1}]_{ii}
		= \sum_{i\in M_g} \dd{}{\sqrt{n}\gamma_{n,g'}} [(\Gamma_n+\bar{U}_n^{\top} H_{E,E} ^{-1} \Lambda_{E} \bar{U}_n)^{-1}]_{ii} \\
		=& \sum_{i\in M_g} \sum_{j,k}
		\dd{[(\Gamma_n+\bar{U}_n^{\top} H_{E,E} ^{-1} \Lambda_{E} \bar{U}_n)^{-1}]_{ii}}{(\Gamma_n+\bar{U}_n^{\top} H_{E,E} ^{-1} \Lambda_{E} \bar{U}_n)_{jk}}
		\dd{(\Gamma_n+\bar{U}_n^{\top} H_{E,E} ^{-1} \Lambda_{E} \bar{U}_n)_{jk}}{\sqrt{n}\gamma_{n,g'}}\\
		=& -\sum_{i\in M_g} \sum_{j,k}
		[(\Gamma_n+\bar{U}_n^{\top} H_{E,E} ^{-1} \Lambda_{E} \bar{U}_n)^{-1}]_{ij}[(\Gamma_n+\bar{U}_n^{\top} H_{E,E} ^{-1} \Lambda_{E} \bar{U}_n)^{-1}]_{ki}
		\dd{(\Gamma_n+\bar{U}_n^{\top} H_{E,E} ^{-1} \Lambda_{E} \bar{U}_n)_{jk}}{\sqrt{n}\gamma_{n,g'}}\\
		=& -\sum_{i\in M_g}\sum_{j\in M_{g'}} [(\Gamma_n+\bar{U}_n^{\top} H_{E,E} ^{-1} \Lambda_{E} \bar{U}_n)^{-1}]_{ij}[(\Gamma_n+\bar{U}_n^{\top} H_{E,E} ^{-1} \Lambda_{E} \bar{U}_n)^{-1}]_{ji}.
	\end{align*}
\end{proof}

\begin{lemma}
Define
	 \begin{equation*}
	I(b, b^\perp, w) = \frac{1}{2}(b- b_E)^\top \Sigma_{E}^{-1}(b- b_E) + \frac{1}{2}(b^\perp- b_{E}^\perp)^\top (\Sigma_{E}^\perp)^{-1}(b^\perp- b_{E}^\perp)+ \frac{1}{2}w^\top \Omega^{-1}w.
         \end{equation*}
Under Assumption 2.2, we have
        the following large deviation limit
        $$
        \lim_{n\rightarrow\infty} \frac{1}{a_n^2} \log \mathbb{P}\left[ \frac{\sqrt{n}}{a_n} \begin{pmatrix} \widehat{\beta}_{n,E} \\ \widehat{\beta}^{\perp}_{n,E} \\ \omega_n \end{pmatrix} \in \mathcal{L}\right]=-   \inf_{(b, b^\perp, w)\in \mathcal{L}} I(b, b^\perp, w)
        $$
        for a convex set $\mathcal{L}\subseteq \mb{R}^{2p}$.
\label{lemma:LDPfull}
\end{lemma}
\begin{proof}
From \cite{AcostaA.De1992}, it follows that $\bar{Z_n}$ satisfies a large deviation limit.
Using the contraction principle for large deviation limits \citep[e.g.,][Section 4.2.1]{DemboZeitouni2009} with the linear transformation $Z'_n = T\bar{Z_n}$, we note that $Z'_n$ also satisfies a large deviation limit.
That is,
$$
        \lim_{n\rightarrow\infty} \frac{1}{a_n^2} \log \mathbb{P}\left[ \frac{\sqrt{n}}{a_n} Z'_n \in \mathcal{L}\right]=-   \inf_{(b, b^\perp, w)\in \mathcal{L}} I(b, b^\perp, w).
        $$
The rate function $I(.)$ is defined in \cite{DemboZeitouni2009} as the Legendre-Fenchel transformation of the logarithmic moment generating function $\log \; M(.)$, which for the limiting Gaussian distribution results in the rate function $I(.)$ and can be calculated as follows. We have the following logarithmic moment generating function
        \begin{align*}
            log \; M_{b, b^\perp, w}(t_1,t_2,t_3)&=\log\left(E\left[e^{t_1^{\top}b_E+\frac{1}{2} t_1^{\top} \Sigma_E t_1} e^{ t_2^{\top}b_E^{\perp}+\frac{1}{2} t_2^{\top} \Sigma_{E^{\perp}} t_2} e^{t_3^{\top}0_p+\frac{1}{2} t_3^{\top} \Omega t_3} \right]\right) \\
            &= t_1^{\top}b_E+\frac{1}{2} t_1^{\top} \Sigma_E t_1 +t_2^{\top}b_E^{\perp}+\frac{1}{2} t_2^{\top} \Sigma_{E^{\perp}} t_2 +\frac{1}{2} t_3^{\top} \Omega t_3
        \end{align*}
        where $t_1 \in \mathbb{R}^{|E|}, t_2 \in \mathbb{R}^{|E'|}$ and $t_3 \in \mathbb{R}^{p}$. The  Legendre-Fenchel transformation is taken
        \begin{align*}
            I(b, b^\perp, w)=\underset{t_1,t_2,t_3}{\max}\big\{ t_1^{\top}b+t_2^{\top}b^\perp+t_3^{\top}w-log \;M_{b, b^\perp, w}(t_1,t_2,t_3)\big\}.
        \end{align*}
        In solving this optimization problem we find $t_1=(b-b_E)^{\top}\Sigma_E^{-1}, t_2=(b^\perp-b^\perp_E)^{\top}(\Sigma_E^\perp)^{-1}$ and $t_3=w^{\top}\Omega^{-1}$. This leads to the rate function
        \begin{align*}
             I(b, b^\perp, w)=&(b-b_E)^{\top}\Sigma_E^{-1}b+(b^\perp-b^\perp_E)^{\top}(\Sigma_E^\perp)^{-1}b^{\perp}+w^{\top}\Omega^{-1}w-((b-b_E)^{\top}\Sigma_E^{-1}b_E\\
             &+\frac{1}{2} (b-b_E)^{\top}\Sigma_E^{-1}(b-b_E)+(b^\perp-b^\perp_E)^{\top}(\Sigma_E^\perp)^{-1}b_E^{\perp}+\frac{1}{2} (b^\perp-b^\perp_E)^{\top}(\Sigma_E^\perp)^{-1}\\ &(b^\perp-b^\perp_E) +\frac{1}{2} w^{\top}\Omega^{-1} w)\\
             =&\frac{1}{2}(b- b_E)^\top \Sigma_{E}^{-1}(b- b_E) + \frac{1}{2}(b^\perp- b_{E}^\perp)^\top (\Sigma_{E}^\perp)^{-1}(b^\perp- b_{E}^\perp)+ \frac{1}{2}w^\top \Omega^{-1}w.
        \end{align*}
Finally, the condition on $\tilde r_n$ in Assumption 2.2 from (10) in the article yield us the limit in the claim.
\end{proof}

 \subsection{Proofs of Main Results}

\subsubsection{Proof of Proposition 3.1}
  \begin{proof}
  Assuming that the entries of $X$ and $\omega_n$ in the set $E$ are stacked before the entries in $E'$, we note that the Group LASSO estimator $\widehat\beta_n^{(\Lambda)}$ satisfies
  \begin{align*}
0_p
	=& \frac{1}{\sqrt{n}} X^\top \nabla \ell (X\widehat\beta_n^{(\Lambda)};Y) +
	\begin{pmatrix}
		\left(\lambda_g \widehat{u}_{n,g}\right)_{g \in \mc{G}_E}\\
		\left(\lambda_g \widehat{z}_{n,g}\right)_{g \in \mc{G}_{E'}}
	\end{pmatrix} -\sqrt{n}\omega_n.
	\end{align*}
	Equivalently,
	\begin{align}
		\sqrt{n}\omega_n &= \frac{1}{\sqrt{n}} X^\top \nabla \ell (X\widehat\beta_n^{(\Lambda)};Y) + C_{\mathcal{E}}\Lambda_{E'} \widehat{z}_n + D_{\mathcal{E}}\widehat{u}_n.
		\label{KKT:stationarity}
	\end{align}
    Denote $\widehat\beta_{n,E}^{(\Lambda)} \in \mathbbm{R}^{|E|}$ the group LASSO estimator restricted to the nonzero rows of the full vector $\widehat\beta_n^{(\Lambda)}$. A Taylor expansion of the first term, in the right-hand side display, around $\widehat{\beta}_{n,E}$ gives us
\begin{eqnarray*}
		\frac{1}{\sqrt{n}} X^\top \nabla \ell (X\widehat\beta_n^{(\Lambda)};Y)
		&=& \frac{1}{\sqrt{n}}X^{\top} \nabla \ell(X_E\widehat{\beta}_{n,E}; Y) + H_{E} \sqrt{n} (\widehat\beta_{n,E}^{(\Lambda)} - \widehat{\beta}_{n,E})  +o_p(1) \\
		&= &H_{E} \sqrt{n} \widehat\beta_{n,E}^{(\Lambda)}
			+\frac{1}{\sqrt{n}}X^{\top} \nabla \ell(X_E\widehat{\beta}_{n,E}; Y)
			- H_{E} \sqrt{n} \widehat{\beta}_{n,E}  + o_p(1)\\
		&= & H_{E} \sqrt{n} \widehat\beta_{n,E}^{(\Lambda)}
			+\frac{1}{\sqrt{n}} \begin{pmatrix} {0}_{|E|} \\
            X_{E'}^{\top} \nabla \ell(X_E\widehat{\beta}_{n,E}; Y) \end{pmatrix}
			- \begin{pmatrix} {0}_{|E|} \\ A_{E}\sqrt{n} \widehat{\beta}_{n,E} \end{pmatrix} \\
    && - K_EH_{E,E}^{-1}\Sigma_{E}^{-1} \sqrt{n} \widehat{\beta}_{n,E}  +o_p(1)\\
		&= &H_{E} \sqrt{n} \widehat{U}_n\widehat{\gamma}_n
			+ C_{\mathcal{E}} \sqrt{n}\widehat{\beta}^{\perp}_{n,E} - K_EH_{E,E}^{-1}\Sigma_{E}^{-1} \sqrt{n} \widehat{\beta}_{n,E}  +o_p(1)\\
		&= & A_{\mathcal{E}} \sqrt{n} \widehat{\beta}_{n,E} + B_{\mathcal{E}}\sqrt{n} \widehat{U}_n\widehat{\gamma}_n + C_{\mathcal{E}} \sqrt{n}\widehat{\beta}^{\perp}_{n,E} +o_p(1).
\end{eqnarray*}
	Plugging this expansion into \eqref{KKT:stationarity} leads us to the representation:
	\begin{align*}
		\sqrt{n}\omega_n
		&=
		A_{\mathcal{E}} \sqrt{n} \widehat{\beta}_{n,E} + B_{\mathcal{E}}\sqrt{n} \widehat{U}_n\widehat{\gamma}_n + C_{\mathcal{E}} (\sqrt{n}\widehat{\beta}^{\perp}_{n,E}+ \Lambda_{E'} \widehat{z}_n) + D_{\mathcal{E}}\widehat{u}_n +o_p(1).
	\end{align*}
\end{proof}

\subsubsection{Proof of Corollary 3.1}
\begin{proof}
From the definition of $\bar{\omega}_n$ and the proof of Proposition 2.1,
$\sqrt{n}\bar{\omega}_n = \sqrt{n}\omega_n + o_p(1)$ and therefore
$\sqrt{n}\bar{\omega}_n \overset{d}{\rightarrow} N_p(0, \Omega)$.
\end{proof}

 \subsubsection{Proof of Proposition 3.2}
 \begin{proof}
 First, consider the mapping
\begin{eqnarray}
	\label{Pi:star}
\Pi^*: \mb{R}^{|E|}\times \mb{R}^{|E'|}\times \mb{R}^{p} \to \mb{R}^{|E|}\times \mb{R}^{|E'|}\times \mb{R}^{p+|E|}:
(b,b^\perp,\omega) \mapsto (b,b^\perp,\Pi^{-1}_{b,b^\perp}(\omega)),
\end{eqnarray}
where we may write $\Pi^{-1}_{b,b^\perp}(\omega)=(g^\top,u^\top,z^\top)^\top$.
Denote the density of
$
\sqrt{n}\big(
 		 \widehat{\beta}^{\top}_{n,E},
 		 (\widehat{\beta}_{n,E}^\perp)^{\top},
 		\bar\omega_n^{\top} \big)^{\top}
$
by $\bar{p}_n$, which exists for $n$ large enough by
Assumption 2.1.
We apply $\Pi^*$ as a change of variables from
$$\sqrt{n}\big(
 		 \widehat{\beta}^{\top}_{n,E},
 		 (\widehat{\beta}_{n,E}^\perp)^{\top},
 		\bar\omega_n^{\top} \big)^{\top} \to \sqrt{n}\big(
 		 \widehat{\beta}^{\top}_{n,E},
 		 (\widehat{\beta}_{n,E}^\perp)^{\top}, \sqrt{n}{\gamma}_n^{\top}, u_n^{\top}, z_n^{\top})^{\top},
$$
and denote the Jacobian matrix associated with $\Pi^*$ by $D_{\Pi^*}(\sqrt{n}\beta_{n,E}, \sqrt{n}\beta_{n,E}^\perp,\sqrt{n}{\gamma}_n, u_n, z_n)$.
Then, we observe that the density of the variables
$
\big(\sqrt{n} \beta_{n,E}^{\top}, \sqrt{n} (\beta_{n,E}^\perp)^{\top}, \sqrt{n} \gamma_n^{\top}, u_n^{\top}, z_n^{\top}\big)^{\top}
$
 is equal to
 \begin{equation}
 \label{den:1}
|\det D_{\Pi^*}(\sqrt{n}\beta_{n,E}, \sqrt{n}\beta_{n,E}^\perp,\sqrt{n}{\gamma}_n, u_n, z_n)|\bar{p}_n(\sqrt{n}\beta_{n,E}, \sqrt{n}\beta_{n,E}^\perp,\Pi_{\sqrt{n}\beta_{n,E}, \sqrt{n}\beta^\perp_{n,E}}(\sqrt{n}\gamma_n, u_n, z_n)).
\end{equation}
Applying the change of variables which undoes the centering
$$
 \sqrt{n}\big(
 		 ({\beta}_{n,E} -b_{n,E}^*)^\top,
 		 ({\beta}_{n,E}^\perp  -b_{n,E}^\perp)^\top,
 		\bar\omega_n^\top \big)^\top
\xmapsto{\mu} \sqrt{n}\big(
 		 {\beta}_{n,E}^\top,
 		 ({\beta}_{n,E}^\perp)^\top,
 		\bar\omega_n^\top\big)
$$
 allows us to write the following
 $$
\bar{p}_n(\sqrt{n}\beta_{n,E}, \sqrt{n}\beta_{n,E}^\perp,\sqrt{n}\bar{w}_n) = p_n(\sqrt{n} ({\beta}_{n,E} - {b}^*_{n,E}), \sqrt{n}({\beta}^{\perp}_{n,E} - {b}_{n,E}^\perp), \sqrt{n}\bar{w}_n).
 $$
 Therefore, the density in \eqref{den:1} is  equal to
$$
|\det D_{\Pi^*}(\sqrt{n}{\gamma}_n, u_n, z_n)|p_n(\sqrt{n}(\beta_{n,E}-b_{n,E}^*), \sqrt{n}(\beta^\perp_{n,E}-b_{n,E}^\perp), \Pi_{\sqrt{n}\beta_{n,E}, \sqrt{n}\beta^\perp_{n,E}}(\sqrt{n}\gamma_n, u_n, z_n)).
$$
To conclude, we use Lemma \ref{lem:Simplified Jacobian of CoV} to note the following equality
 \begin{align*}
 	|\det D_{\Pi^*}(\sqrt{n}\beta_{n,E}, \sqrt{n}\beta_{n,E}^\perp,\sqrt{n}{\gamma}_n, u_n, z_n) |
 	=
 	|\det D_{\Pi}(\sqrt{n}{\gamma}_n, u_n, z_n)|,
 \end{align*}
which completes the proof.
 \end{proof}

 \subsubsection{Proof of Proposition 3.3}
 \begin{proof}
We start from the following asymptotic representation obtained by taking a Taylor series expansion of the gradient of the loss around $b_{n,E}^*$,
\begin{equation}
\label{orth:est}
\begin{aligned}
\frac{1}{\sqrt{n}}X^{\top} \nabla \ell(X_E\widehat{\beta}_{n,E}; Y) &= \frac{1}{\sqrt{n}}X^{\top} \nabla \ell(X_E b_{n,E}^*; Y) + H_{E} \sqrt{n} (\widehat{\beta}_{n,E} - b_{n,E}^*)  +o_p(1).
\end{aligned}
\end{equation}
We consider separately the first $|E|$ rows of this vector and the last $|E'|$ rows.
By definition of the (refitted) estimator in the selected model,
\begin{equation*}
\frac{1}{\sqrt{n}}X_E^{\top} \nabla \ell(X_E\widehat{\beta}_{n,E}; Y) = 0_{|E|}
=  \frac{1}{\sqrt{n}}X_E^{\top} \nabla \ell(X_E b_{n,E}^*; Y) + H_{E, E} \sqrt{n} (\widehat{\beta}_{n,E} - b_{n,E}^*)  +o_p(1).
\end{equation*}
That is, we have
\begin{equation}
\label{M:est}
\sqrt{n} (\widehat{\beta}_{n,E} - b_{n,E}^*) =  -H_{E, E}^{-1}\frac{1}{\sqrt{n}}X_E^{\top} \nabla \ell(X_E b_{n,E}^*; Y)  +o_p(1).
\end{equation}
Using the definition of $\Sigma_E$ which corresponds to a `sandwich' covariance structure, it follows that
$
 \sqrt{n} (\widehat{\beta}_{n,E} - b_{n,E}^*)\stackrel{d}{\to} \mathcal{N}_p(0, \Sigma_E).
$

Next, we start from the definition of $\widehat{\beta}^{\perp}_{n,E}$ and we take a Taylor expansion of the last $|E'|$ rows of the gradient vector around $b_{n,E}^*$,
\begin{equation}
\begin{aligned}
	\label{eq:betaperp}
	\sqrt{n}\widehat{\beta}_{n,E}^{\perp}
	=& \frac{1}{\sqrt{n}}X_{E'}^{\top}
	\nabla \ell(X_E b_{n,E}^*; Y)
	+ H_{E',E} \sqrt{n} (\widehat{\beta}_{n,E} - b_{n,E}^*)\\
	&- A_{E}\sqrt{n} \widehat{\beta}_{n,E}
	+ o_p(1).
\end{aligned}
\end{equation}
With $V_{n,E'} = \mathbb{E}[n^{-1}X_{E'}^{\top}\nabla \ell(X_E b_{n,E}^*; Y)]$ and the definition of $b_{n,E}^\perp$,
we have
\begin{eqnarray*}
\sqrt{n}(\widehat{\beta}_{n,E}^{\perp} -  b_{n,E}^\perp)
&=&
	\big(\frac{1}{\sqrt{n}}X_{E'}^{\top}
	\nabla \ell(X_E b_{n,E}^*; Y)- \sqrt{n}
		V_{n,E'}\big)
\\
&&
	+ H_{E',E} \sqrt{n} (\widehat{\beta}_{n,E} - b_{n,E}^*)
	- A_E\sqrt{n} (\widehat{\beta}_{n,E} - b_{n,E}^*)
	+ o_p(1).
\end{eqnarray*}
Since
\begin{align*}
	&\textup{Cov}\left(
	\frac{1}{\sqrt{n}}X_{E'}^{\top} \nabla \ell(X_E b_{n,E}^*; Y),
	-(H_{E',E} - A_{E}) H_{E,E}^{-1} \frac{1}{\sqrt{n}}X_E^{\top} \nabla \ell(X_E b_{n,E}^*; Y)\right) \\
	=& -K_{E',E} H_{E,E}^{-1} (H_{E,E'} - A_E^\top) = -K_{E',E} K_{E,E}^{-1} K_{E, E'},
\end{align*}
we have
\begin{align*}
	\Sigma^{\perp}_E = K_{E',E'} - 2K_{E',E} K_{E,E}^{-1} K_{E, E'} + K_{E',E} K_{E,E}^{-1} K_{E, E'} = K_{E',E'}  - K_{E',E} K_{E,E}^{-1} K_{E, E'}
\end{align*}
and it follows that $\sqrt{n}(\widehat{\beta}^{\perp}_{n,E} - {b}_{n,E}^\perp) \stackrel{d}{\to}
 	\mathcal{N}_{|E'|}(0_{|E'|}, \Sigma^{\perp}_E),$ as $n\to\infty$.
Using the asymptotic representations in \eqref{orth:est} and \eqref{M:est}, we further observe that the asymptotic covariance between $\widehat{\beta}_{n,E}$ and $\widehat{\beta}_{n,E}^{\perp}$ is $0$.

Finally, the asymptotic independence between both $\sqrt{n}\widehat{\beta}_{n,E}, \sqrt{n}\widehat{\beta}_{n,E}^{\perp}$ and $\sqrt{n}\bar{\omega}_n$ follows from the independence between $\sqrt{n}\widehat{\beta}_{n,E}, \sqrt{n}\widehat{\beta}_{n,E}^{\perp}$ and $\sqrt{n}\omega_n$, and the fact that $\sqrt{n}\bar{\omega}_n = \sqrt{n}{\omega}_n + o_p(1)$.
\end{proof}

\subsubsection{Proof of Proposition 3.4}
 \begin{proof}
Under Assumption 2.0 from which follows that $p_n$ converges to a limiting Gaussian density, the `asymptotic' density of
$ \big(\sqrt{n} \widehat{\beta}_{n,E}^\top,
			\sqrt{n} (\widehat{\beta}^{\perp}_{n,E})^\top,
			\widehat{\gamma}_{n}^\top,
			 \widehat{u}_n^\top,
			  \widehat{z}_n^\top\big)^\top
$
		is equal to
\begin{equation}
		\begin{aligned}
		\label{bigjointgaussian}
		f_n({\beta}_{n,E}, {\beta}_{n,E}^\perp, \gamma, u, z) =
		&\phi(\sqrt{n} {\beta}_{n,E}; \sqrt{n} b_{n,E}^*, \Sigma_E)
		\phi(\sqrt{n} {\beta}_{n,E}^{\perp}; \sqrt{n} {b}_{n,E}^\perp, \Sigma_E^\perp)
		 \\
		&\times \phi\left(\Pi_{\sqrt{n}{\beta}_{n,E}, \sqrt{n}{\beta}_{n,E}^{\perp}}(\sqrt{n}\gamma, u, z); 0, \Omega\right) |\det D_{\Pi}(\sqrt{n}\gamma, u, z)|.
	\end{aligned}
	\end{equation}
When we further condition on
$
\{\widehat{\beta}^{\perp}_{n,E}=  \beta^{\perp}_{n,E}, \widehat{u}_n= u_n, \ \widehat{z}_n = z_n\},
$
and use  Lemma \ref{lem:JacobianComputation} to get
	$|\det D_{\Pi}(\sqrt{n}\gamma_n, u_n, z_n)| = J_{\Pi}(\sqrt{n}\gamma_n, u_n)
$,
we obtain the following density in the limit
	\begin{equation}
	\small
	\begin{aligned}
		\label{eqn:impliedGaussian}
		& {f_n({\beta}_{n,E}, {\beta}_{n,E}^\perp, \gamma_n, u_n, z_n)}\big/
		{\int f_n(b, {\beta}_{n,E}^\perp, g, u_n, z_n)\ db\ dg} \\
		%
		= &
		 \frac{\phi(\sqrt{n} {\beta}_{n,E}; \sqrt{n} b_{n,E}^*, \Sigma_E)
		\phi\left(\Pi_{\sqrt{n}{\beta}_{n,E}, \sqrt{n}{\beta}_{n,E}^{\perp}}(\sqrt{n}\gamma_n, u_n, z_n); 0, \Omega\right)
		J_{\Pi}(\sqrt{n}\gamma_n, u_n)}{\int \phi(\sqrt{n} b; \sqrt{n} b_{n,E}^*, \Sigma_E)
		\phi\left(\Pi_{\sqrt{n}b, \sqrt{n}{\beta}_{n,E}^{\perp}}(\sqrt{n}g, u_n, z_n); 0, \Omega\right) J_{\Pi}(\sqrt{n}g, u_n)\ db\ dg}.
	\end{aligned}
	\end{equation}
		We can further simplify this to obtain, with $c_E=1/\{(2\pi)^{|E|+p}|\Sigma_E|\cdot|\Omega|\}^{1/2}$,
    \begin{align*}
    &\phi(\sqrt{n} {\beta}_{n,E}; \sqrt{n} b_{n,E}^*, \Sigma_E)
		\phi\left(\Pi_{\sqrt{n}{\beta}_{n,E}, \sqrt{n}{\beta}_{n,E}^{\perp}}(\sqrt{n}\gamma_n, u_n, z_n); 0, \Omega\right) \\
   =&c_E\exp\big\{-\frac{1}{2}(\sqrt{n} \beta_{n,E}-\sqrt{n} b_{n,E}^*)^{\top} \Sigma_E^{-1} (\sqrt{n} \beta_{n,E}-\sqrt{n} b_{n,E}^*)\big\} \\
    &\times \exp\big\{\frac{1}{2}(A_{\mathcal{E}}\sqrt{n}\beta_{n,E}+B_{\mathcal{E}}\widehat{U}_n \sqrt{n} \gamma_n+C_{\mathcal{E}} \Lambda_{E'} z_n +D_{\mathcal{E}}u_n+C_{\mathcal{E}}\sqrt{n}\beta^{\perp}_{n,E})^{\top} \\
    &\times \Omega^{-1} (A_{\mathcal{E}}\sqrt{n}\beta_{n,E}+B_{\mathcal{E}}\widehat{U}_n \sqrt{n} \gamma_n+C_{\mathcal{E}} \Lambda_{E'} z_n +D_{\mathcal{E}}u_n+C_{\mathcal{E}}\sqrt{n}\beta^{\perp}_{n,E})\big\} \\
    =& c_E \exp\big\{-\frac{1}{2}\big[n \beta^{\top}_{n,E} \Sigma^{-1}_E \beta_{n,E}-2n\beta^{\top}_{n,E}\Sigma^{-1}_E b_{n,E}^*+n(b_{n,E}^*)^{\top} \Sigma^{-1}_E b_{n,E}^* \\
        &+ 2 \sqrt{n} \gamma_n^{\top}(B_{\mathcal{E}}\widehat{U}_n)^{\top}\Omega^{-1}(A_{\mathcal{E}} \sqrt{n} \beta_{n,E} +C_{\mathcal{E}}\Lambda_{E'}z_n+D_{\mathcal{E}}u_n+C_{\mathcal{E}}\sqrt{n} \beta_{n,E}^{\perp}) \\
        &+ (A_{\mathcal{E}} \sqrt{n} \beta_{n,E} +C_{\mathcal{E}}\Lambda_{E'}z_n+D_{\mathcal{E}}u_n+C_{\mathcal{E}}\sqrt{n} \beta_{n,E}^{\perp})^{\top} \Omega^{-1} (A_{\mathcal{E}} \sqrt{n} \beta_{n,E} +C_{\mathcal{E}}\Lambda_{E'}z_n+D_{\mathcal{E}}u_n+C_{\mathcal{E}}\sqrt{n} \beta_{n,E}^{\perp}) \\
        &+ n \gamma_n^{\top} (B_{\mathcal{E}}\widehat{U}_n)^{\top}\Omega^{-1}B_{\mathcal{E}}\widehat{U}_n \gamma_n\big]\big\}\\
    =& c_E \exp\big\{-\frac{1}{2}n \beta^{\top}_{n,E} \Sigma^{-1}_E \beta_{n,E}+n\beta^{\top}_{n,E}\Sigma^{-1}_E b_{n,E}^*-\frac{1}{2}n(b_{n,E}^*)^{\top} \Sigma^{-1}_E b_{n,E}^*\\
        &-\frac{1}{2}(\sqrt{n} \gamma_n-\bar{A}\sqrt{n}\beta_{n,E}-\bar{b})^{\top} \bar{\Omega}^{-1}(\sqrt{n} \gamma_n-\bar{A}\sqrt{n}\beta_{n,E}-\bar{b})+\frac{1}{2}[\sqrt{n}\beta_{n,E}^{\top}\bar{A}^{\top}\bar{\Omega}^{-1} \bar{A}\beta_{n,E}\sqrt{n}\\
        &+2\sqrt{n}\beta_{n,E}^{\top}\bar{\Omega}^{-1}\bar{\Omega}\bar{A}^{\top}\bar{\Omega}^{-1}\bar{b}+\bar{b}^{\top}\bar{\Omega}^{-1}\bar{b}]
        -\frac{1}{2}[n\beta_{n,E}^{\top}A_{\mathcal{E}}^{\top}\Omega^{-1}A_{\mathcal{E}}\beta_{n,E}
        +2\sqrt{n}\beta_{n,E}^{\top}\bar{\Omega}^{-1}\bar{\Omega}A_{\mathcal{E}}^{\top}\Omega^{-1} \\
        &\times (C_{\mathcal{E}}\Lambda_{E'}z_n+D_{\mathcal{E}}u_n+C_{\mathcal{E}}\sqrt{n} \beta_{n,E}^{\perp})+(C_{\mathcal{E}}\Lambda_{E'}z_n+D_{\mathcal{E}}u_n+C_{\mathcal{E}}\sqrt{n} \beta_{n,E}^{\perp})^{\top}\Omega^{-1} \\&(C_{\mathcal{E}}\Lambda_{E'}z_n+D_{\mathcal{E}}u_n+C_{\mathcal{E}}\sqrt{n} \beta_{n,E}^{\perp})]\big\}\\
        =& c_E \exp\big\{(-\frac{1}{2} n \beta_{n,E}^{\top}\bar{\Theta}^{-1}\beta_{n,E}+\sqrt{n}\beta_{n,E}^{\top} \bar{\Theta}^{-1} \bar{s} + n \beta_{n,E}^{\top} \bar{\Theta}^{-1} \bar{R} b_{n,E}^*\\
        &+c(\sqrt{n}b_{n,E}^*)-
        \frac{1}{2}(\sqrt{n}\gamma_n-\bar{A}\sqrt{n}\beta_{n,E}-\bar{b})^{\top}\bar{\Omega}^{-1}(\sqrt{n}\gamma_n-\bar{A}\sqrt{n}\beta_{n,E}-\bar{b})\big\} \\
        =& c_E \exp\big\{-\frac{1}{2}\big[(\sqrt{n}\beta_{n,E}-\bar{R}\sqrt{n} b^{*}_{n,E}-\bar{s})^{\top} (\bar{\Theta})^{-1} (\sqrt{n}\beta_{n,E}-\bar{R}\sqrt{n} b^{*}_{n,E}-\bar{s})\\
        &-(\bar{R}\sqrt{n}b_{n,E}^*+(\bar{s}))^{\top}\bar{\Theta}^{-1}(\bar{R}\sqrt{n}b_{n,E}^*+\bar{s}) +c(\sqrt{n}b_{n,E}^*) \\
        &-\frac{1}{2}(\sqrt{n}\gamma_n-\bar{A}\sqrt{n}{\beta}_{n,E}-\bar{b})^{\top} \bar{\Omega}^{-1} (\sqrt{n}\gamma_n-\bar{A}\sqrt{n}{\beta}_{n,E}-\bar{b})\big]\big\}
        \\
        =& {k}(\sqrt{n} b_{n,E}^*)  \phi(\sqrt{n}{\beta}_{n,E};         \bar{R}\sqrt{n} b^{*}_{n,E}+ \bar{s}, \bar{\Theta})         \phi(\sqrt{n}{\gamma}_n; \bar{A}\sqrt{n}{\beta}_{n,E}+ \bar{b}, \bar{\Omega}),
\end{align*}
 where we used the notation defined above Proposition 2.4 and in addition defined
\begin{eqnarray*}
c(\sqrt{n}b_{n,E}^*) & = &
\bar{b}^{\top} \bar{\Omega}^{-1}\bar{b} -\frac{1}{2}(C_{\mathcal{E}} \Lambda_{E'} z_n+D_{\mathcal{E}}u_n+C_{\mathcal{E}} \sqrt{n}\beta^{\perp}_{n,E})^{\top} \Omega^{-1} \\
&& \times(C_{\mathcal{E}} \Lambda_{E'} z_n+D_{\mathcal{E}}u_n+C_{\mathcal{E}} \sqrt{n}\beta^{\perp}_{n,E}) -\frac{1}{2}n (b_{n,E}^{*})^{\top}\Sigma^{-1}_E b_{n,E}^{*}, \\
{k}(\sqrt{n} b_{n,E}^*) & = & c_E\exp\{-( \bar{R}\sqrt{n}b_{n,E}^*+\bar{s})^{\top}\bar{\Theta}^{-1}(\bar{R}\sqrt{n}b_{n,E}^*+\bar{s})+c(\sqrt{n}b_{n,E}^*)\},
\end{eqnarray*}
which are both free of $\beta_{n,E}$ and $\gamma_n$, as is the denominator in \eqref{eqn:impliedGaussian}.
Thus, the numerator in \eqref{eqn:impliedGaussian} can be written as
$
	J_{\Pi}(\sqrt{n}\widehat{\gamma}_n, u_n) \phi(\sqrt{n}\widehat{\beta}_{n,E}; \bar{R}\sqrt{n} b^{*}_{n,E}+ \bar{s}, \bar{\Theta})  \phi(\sqrt{n}\widehat{\gamma}_n; \bar{A}\sqrt{n}\widehat{\beta}_{n,E}+ \bar{b}, \bar{\Omega})
	$ 
and the statement is proven.
\end{proof}

\subsubsection{Proof of Theorem 3.2}
\begin{proof}
	Fixing some notations for our proof, let
	  $S_n = \big(\frac{\sqrt{n}}{a_n}\widehat\beta_{n,E}^\top, \frac{\sqrt{n}}{a_n}\widehat(\beta^\perp_{n,E})^\top,
     \frac{\sqrt{n}}{a_n}\widehat\gamma_{n}^\top, \widehat{U}_n^\top, \frac{1}{a_n}\widehat{Z}_n^\top\big)^\top$,
      and
	 $T_n = \big( \widehat{\beta}_{n,E}^\top, (\widehat{\beta}^{\perp}_{n,E})^\top, \widehat{V}_n^\top \big)^\top$
	 where
	 $
	 \sqrt{n}\widehat{V}_n = \sqrt{n}\bar\omega_n - D_{\mathcal{E}} \widehat{U}_n,
	 $
	 and $\|\widehat{U}_n\|_2 =1$.	
	 Define
         $$
        \Delta_{b, b^\perp}(g, u, z)= \Pi_{b, b^\perp}(g, u, z)- D_{\mathcal{E}}u = A_{\mathcal{E}}b + B_{\mc{E}}\text{bd}(u) g + C_{\mathcal{E}}(\Lambda_{E'} z + b^{\perp}).
         $$
          Due to Assumption 2.3 and the large deviation limit in Lemma \ref{lemma:LDPfull}, where we take $C_n = - D_{\mathcal{E}} \widehat{U}_n$, we note that
        $$\lim_{n\rightarrow\infty} \frac{1}{a_n^2} \log \mathbb{P}\left[ \frac{\sqrt{n}}{a_n}T_n \in \mathcal{L}\right]=-\inf_{(b, b^\perp, w)\in \mathcal{L}} I(b, b^\perp, w),$$
         because
         $$
         \lim_{n\rightarrow\infty} \left\{ \frac{1}{a_n^2} \log \mathbb{P}\left[ \frac{\sqrt{n}}{a_n}\big(\widehat{\beta}_{n,E}^\top, (\widehat{\beta}^{\perp}_{n,E})^\top,
          \bar\omega_n^\top \big) \in \mathcal{L}\right] -\frac{1}{a_n^2} \log \mathbb{P}\left[ \frac{\sqrt{n}}{a_n}T_n \in \mathcal{L}\right] \right\} =0.
         $$

         We now apply two steps to prove the limit in our claim.
	In the first step, we consider the mapping
	$
    \psi: (b, b^\perp, v) \to (b, b^\perp, g, u, z)
    $
     such that
	\begin{align*}
		\psi^{-1}\big(b, b^\perp, g, u, z\big)
		 = \big( b^\top, (b^\perp)^\top, \Delta_{b, b^\perp}(g, u, z)^\top \big)^\top=
 \big( b^\top,  (b^\perp)^\top, v^\top \big)^\top.
	\end{align*}
        Based on $\Pi_{b, b^{\perp}}$, defined as per Proposition 2.1, and the definition of $\widehat{V}_n$, note that
         $$\frac{\sqrt{n}}{a_n}\widehat{V}_n = \Delta_{\frac{\sqrt{n}}{a_n}\widehat\beta_{n,E}, \frac{\sqrt{n}}{a_n}\widehat\beta^\perp_{n,E}}\left(\frac{\sqrt{n}}{a_n}\widehat{\gamma}_n, \widehat{u}_n, \frac{1}{a_n}\widehat{z}_n\right),$$
         and thus,
         $
         S_n = \psi ({\sqrt{n}}/{a_n} T_n).
         $

         Therefore, applying the contraction principle for large deviation limits, we observe that
	$$\lim_{n\rightarrow\infty} \frac{1}{a_n^2} \log \mathbb{P}\left[ S_n \in \mathcal{L}_0\right]=-\inf_{(b, b^\perp, g, u, z)\in \mathcal{L}_0} I\circ \psi^{-1}(b, b^\perp, g, u, z),$$
	 for a convex set $\mathcal{L}_0$.
	In particular, the optimization objective on the right-hand side display is equal to
	\begin{equation*}
	\begin{aligned}
	I\circ \psi^{-1}(b, b^\perp, g, u, z) &= \frac{1}{2}(b- b_E)^\top \Sigma_{E}^{-1}(b- b_E) + \frac{1}{2}(b^\perp- b_{E}^\perp)^\top (\Sigma_{E}^\perp)^{-1}(b^\perp- b_{E}^\perp)\\
	&\;\; + \frac{1}{2}(\Delta_{b, b^{\perp}}(g, u, z))^\top \Omega^{-1}\Delta_{b, b^{\perp}}(g, u, z).
	\end{aligned}
	\end{equation*}

In the second step, we use the large deviation principle for the conditional probability as in \citet{LaCourSchieve2015} and again the contraction principle
to arrive at
	\begin{equation*}
	\begin{aligned}
	&\lim_{n\rightarrow\infty} \frac{1}{a_n^2} \log \mathbb{P}\bigg[ \frac{\sqrt{n}}{a_n}\widehat\gamma_{n}\in \mathcal{K}\   \bigg{|}\
		\frac{\sqrt{n}}{a_n}\widehat\beta_{n,E}^{\perp} = \beta_{n,E}^\perp, \widehat{u}_n = u_n, \frac{1}{a_n}\widehat{z}_n = z_n\bigg] \\
        &= -\inf_{b, (g \in \mathcal{K})} \left\{I\circ \psi^{-1}(b, \beta_{n,E}^{\perp}, g, u_n, z_n) - \inf_{b_0, g_0}I\circ \psi^{-1}(b_0, \beta_{n,E}^{\perp}, g_0, u_n, z_n)\right\}\\
        &= - \inf_{b, (g \in \mathcal{K})} \bigg\{ \frac{1}{2}(b - \bar{R}b_E - \bar{s}_0)^\top(\bar{\Theta})^{-1}(b - \bar{R}b_E - \bar{s}_0)
		+\frac{1}{2}(g - \bar{A}b - \bar{b}_0)^\top\bar{\Omega}^{-1}(g - \bar{A}b- \bar{b}_0)\bigg\},
	\end{aligned}
	\end{equation*}
   where
   $
   a_n \bar{b}_0 = \bar{b}+\bar\Omega (B_{\mathcal{E}}bd(u_n))^\top \Omega^{-1} D_{\mathcal{E}}u_n
   $
   and
   $a_n \bar{s}_0 = \bar{s} - \bar{\Theta}A_{\mathcal{E}}^\top \Omega^{-1} (B_{\mathcal{E}}bd(u_n)\bar{\Omega}(B_{\mathcal{E}}bd(u_n))^{\top}$
   $\times \Omega^{-1}-I)D_{\mathcal{E}}u_n$.
We conclude the proof by observing that for the constrained optimization such that the $\widehat\gamma_{n}$ are strictly positive,
\begin{equation*}
\begin{aligned}
& \lim_{n\to \infty}  \inf_{b', g' } \bigg\{ \frac{1}{2}\left(b' - \bar{R} b_E - \frac{1}{a_n}\bar{s}\right)^\top\bar{\Theta}^{-1}\left(b' - \bar{R} b_E - \frac{1}{a_n}\bar{s}\right)\\
&+ \frac{1}{2}\left(g' - \bar{A}b' - \frac{1}{a_n}\bar{b}\right)^\top\bar{\Omega}^{-1}\left(g' - \bar{A}b' - \frac{1}{a_n}\bar{b}\right) - \frac{1}{a_n^2}\log J(a_ng'; U_n)+\frac{1}{a_n^2}{\text{\normalfont Barr}}_{\mathcal{K}}(a_n g')\bigg\}   \\
&= \inf_{b, (g \in \mathcal{K})} \bigg\{ \frac{1}{2}(b - \bar{R}b_E - \bar{s}_0)^\top\bar{\Theta}^{-1}(b - \bar{R}b_E - \bar{s}_0)
		+\frac{1}{2}(g - \bar{A}b - \bar{b}_0)^\top\bar{\Omega}^{-1}(g - \bar{A}b- \bar{b}_0)\bigg\}.
\end{aligned}
\end{equation*}
The above stated conclusion follows by noting that the Jacobian satisfies
$ \sup_{g \in \mathcal{K}} J(a_ng'; U_n)= O(a_n),$
and that the sequence of convex objectives in the left-hand side display converge to the convex objective on the right-hand side display, and that the optimization on the right-hand side display has a unique minimum. See Theorem 2.2 in \cite{AcostaA.De1992} for the condition $a_n=o(n^{1/2})$.
\end{proof}

\subsubsection{Proof of Theorem 4.1}
\begin{proof}
	In finding the MLE, we reparameterize the log-likelihood
	\begin{equation}
 \begin{aligned}
 	& \log \phi(\sqrt{n}\widehat{\beta}_{n,E}; \bar{R}\sqrt{n} b^{*}_{n,E}+ \bar{s}, \bar{\Theta})  + \inf_{b, g}
		\bigg\{ \frac{1}{2}\left(\sqrt{n} b - \bar{R}\sqrt{n} b^*_{n,E} - \bar{s}\right)^\top\bar{\Theta}^{-1}\left(\sqrt{n} b - \bar{R}\sqrt{n}b^*_{n,E} - \bar{s}\right)\\
		&\;+ \frac{1}{2}\left(\sqrt{n} g - \bar{A}\sqrt{n} b - \bar{b}\right)^\top\bar{\Omega}^{-1}\left(\sqrt{n} g - \bar{A}\sqrt{n} b - \bar{b}\right) - \log J(\sqrt{n}g; U_n)+{\text{\normalfont Barr}}_{\mathcal{K}}(\sqrt{n}g)\bigg\}.
 \end{aligned}
 \label{asymptotic:pslik}
\end{equation}
	in terms of
	\begin{align} \label{eq:def_eta}
		\eta_{n,E} = \bar{\Theta}^{-1}\left(\bar{R}\sqrt{n}b^*_{n,E} +\bar{s}\right).
	\end{align}
We define the function $h:\mb{R}^{|E|}\times \mb{R}^{|E|} \to \mb{R}$ such that
	\begin{eqnarray*}
		h(\widetilde{b}, \widetilde{g}) =
		\frac{1}{2} \widetilde{b}^\top\bar{\Theta}^{-1}\widetilde{b}
		+ \frac{1}{2}(\widetilde{g} - \bar{A}\widetilde{b} - \bar{b})^\top\bar{\Omega}^{-1}(\widetilde{g} - \bar{A}\widetilde{b} - \bar{b})
		- \log J_{\Pi}(\widetilde{g}, u_n)+{\text{\normalfont Barr}}_{\mathcal{K}}(\widetilde{g}).
	\end{eqnarray*}
The convex conjugate function of $h$ at $(\eta_{n,E}^\top, 0_{|E|}^\top)^\top$ is
$
		h^*(\eta_{n,E}, 0_{|E|}) = \sup_{\widetilde{b}, \widetilde{g}}
		\big\{ \widetilde{b}^\top \eta_{n,E} - h(\widetilde{b},\widetilde{g}) \big\}.
$

	Now, we turn to our approximate post-selection log-likelihood in \eqref{asymptotic:pslik}, which is equal to
	\begin{align} \label{eq:approx_loglik}
		-\frac{1}{2}n \widehat{\beta}_{n,E}^\top(\bar{\Theta})^{-1} \widehat{\beta}_{n,E} + \sqrt{n}\widehat{\beta}_{n,E}^\top\eta_{n,E}  - h^*(\eta_{n,E}, 0_{|E|}).
	\end{align}
	It follows that we can rewrite the MLE as
	$
    \sqrt{n}\widehat{b}^{mle}_{n,E} = \bar{R}^{-1}\bar{\Theta} \eta_{n,E}^* - \bar{R}^{-1} \bar{s},
    $
    with the current parametrization
	\begin{align}
		\eta_{n,E}^* = \underset{\eta_{n,E}}{\arg\min}\big\{ h^*(\eta_{n,E}, 0_{|E|}) -
	\sqrt{n}\widehat{\beta}_{n,E}^\top\eta_{n,E} \big\}.
	\label{eta:problem}
	\end{align}
	Next, to minimize over $\eta_{n,E}$, we notice that \eqref{eta:problem} is equivalently written as the following constrained optimization problem
	\begin{align*}
		&\underset{\eta_{n,E},\eta_{n,E}',u}{\min}\big\{\ h^*(\eta_{n,E}', u) -
	\sqrt{n}\widehat{\beta}_{n,E}^\top\eta_{n,E} \big\} \quad \text{such that} \quad \eta_{n,E} = \eta_{n,E}', u = 0_{|E|}.
	\end{align*}
	The dual problem with Lagrange multipliers $b$ and $g$ is equal to
$ 
		\underset{b,g}{\max} \underset{\eta_{n,E},\eta_{n,E}',u}{\min} \mathcal{L}(b,g;\eta_{n,E},\eta_{n,E}',u),
$ 
	with the Lagrangian
	\begin{align*}
		\mathcal{L}(b,g;\eta_{n,E},\eta_{n,E}',u)
		&= b^\top(\eta_{n,E} -\eta_{n,E}') - g^\top u + h^*(\eta_{n,E}', u) -
		\sqrt{n}\widehat{\beta}_{n,E}^\top\eta_{n,E}\\
		&= \left[-b^\top \eta_{n,E}' - g^\top u + h^*(\eta_{n,E}', u) \right] +
		\left[b^\top\eta_{n,E} -
		\sqrt{n}\widehat{\beta}_{n,E}^\top\eta_{n,E} \right].
	\end{align*}
Observing the definition of the conjugate function of $h^*$, and the convexity of $h$,
$$
\underset{\eta_{n,E}',u}{\min}\{ -b^\top \eta_{n,E}' - g^\top u + h^*(\eta_{n,E}', u) \}=-h^{**}(b,g) = -h(b,g).
$$
Hence, by definition of a dual function, and switching the orders of the minimum and maximum,
	\begin{align}
&\underset{b,g}{\max}\underset{\eta_{n,E},\eta_{n,E}',u}{\min}\big[ \mathcal{L}(b,g;\eta_{n,E},\eta_{n,E}',u)  \big]
= \underset{b,g}{\max}\min_{\eta_{n,E}}
\big\{ -h(b,g)+  (b - \sqrt{n}\widehat{\beta}_{n,E})^\top\eta_{n,E} \big\}  \nonumber \\
& =  \underset{\eta_{n,E}}{\min}\big[ \underset{b,g}{\max} \big\{ -h(b,g)+  (b - \sqrt{n}\widehat{\beta}_{n,E})^\top\eta_{n,E} \big\}  \big].
\label{eq:opt_h_eta}
\end{align}
We recognize here a constrained optimization problem:
$\min_{b,g} h(b,g)$ such that $b=\sqrt{n}\widehat\beta_{n,E}$, using $\eta_{n,E}$ as the Lagrange multiplier.
Hence, \eqref{eq:opt_h_eta} is equal to $\min_g h(\sqrt{n}\widehat\beta_{n,E},g)$,
 which reduces the original problem to an optimization over $g\in \mb{R}^{|E|}$.

	
	Denote the solution to the dual problem by
	\begin{align*}
		(b^*, g^*, \eta_{n,E}^*,\eta_{n,E}'^*,u^*) = \arg\ \underset{b,g}{\max}\underset{\eta_{n,E},\eta_{n,E}',u}{\min} \mathcal{L}(b,g;\eta_{n,E},\eta_{n,E}',u).
	\end{align*}
	Because our constrained optimization problem is convex and strictly feasible, the KKT stationarity conditions give us
    $
    \nabla h^*(\eta_{n,E}'^*,u^*) = (b^{*\top}, g^{*\top})^\top,
	$
	and by \eqref{eq:opt_h_eta}, we know
	$b^* = \sqrt{n}\widehat{\beta}_{n,E}$,
		$g^* = \arg \min_g \ h(\sqrt{n}\widehat{\beta}_{n,E},g)$.
	Therefore,
	\begin{align} \label{eq:etaprime_u_nablah}
		(\eta_{n,E}'^*,u^*) = (\nabla h^*)^{-1}(b^*, g^*) = \nabla h(b^*, g^*) = \nabla h(\sqrt{n}\widehat{\beta}_{n,E}, g^*),
	\end{align}
	which leads us to
$ 
		\eta_{n,E}'^* = (\bar{\Theta})^{-1}\sqrt{n}\widehat{\beta}_{n,E} + \bar{A}^\top\bar{\Omega}^{-1} \bar{A}\sqrt{n}\widehat{\beta}_{n,E} + \bar{b} - g^*).
	$ 
Defining $ g_n^*(\sqrt{n}\widehat{\beta}_{n,E}) = g^*$
	and using the constraint $\eta_{n,E}^* = \eta_{n,E}'^*$, we arrive at the final expression for the MLE
	\begin{align*}
		\sqrt{n}\widehat{b}^{mle}_{n,E}
		&= \bar{R}^{-1}\sqrt{n}\widehat{\beta}_{n,E} - \bar{R}^{-1} \bar{s} + \Sigma_E \bar{A}^\top\bar{\Omega}^{-1}\left(\bar{A}\sqrt{n}\widehat{\beta}_{n,E} + \bar{b} - g_n^*(\sqrt{n}\widehat{\beta}_{n,E})\right).
	\end{align*}

We continue with the calculation of the observed Fisher information matrix.

Using \eqref{eta:problem}, for a given $\eta_{n,E} = \eta_{n,E}(b^*_{n,E}) = \bar{\Theta}^{-1}\left(\bar{R}\sqrt{n}b^*_{n,E} +\bar{s}\right)$, define
	\begin{eqnarray} \label{eq:b_and_gtilde}
		\begin{pmatrix}
			\widetilde{b}^*\\
			\widetilde{g}^*
		\end{pmatrix}
		\defeq
		(\nabla h)^{-1}(\eta_{n,E},0_{|E|})
        =\nabla h^*(\eta_{n,E},0_{|E|})
		=
		\begin{pmatrix}
			\dd{}{\eta}h^*(\eta, \lambda)\\
			\dd{}{\lambda}h^*(\eta, \lambda)
		\end{pmatrix}
		\bigg|_{(\eta_{n,E},0_{|E|})} .
	\end{eqnarray}
	Note that $\widetilde{b}^*, \widetilde{g}^*$ are functions of $\eta_{n,E}$.
	Ignoring constants, our approximate log-likelihood \eqref{eq:approx_loglik} agrees with
$ 
		f(\eta_{n,E}) =\sqrt{n}\widehat{\beta}_{n,E}^\top\eta_{n,E} -
		h^*(\eta_{n,E}, 0_{|E|}).
$ 
       Using the chain rule, and the fact that from \eqref{eq:def_eta},
    $	
    \dd{\eta_{n,E}}{b^*_{n,E}} = \sqrt{n}\bar{\Theta}^{-1} \bar{R} = \sqrt{n}\Sigma_E^{-1},
	$
	the observed Fisher Information matrix in $b^*_{n,E}$ is equal to
	\begin{align*}
		\dd{^2}{{b^{*2}_{n,E}}} \left[-f(\eta_{n,E})\right]
		=
		n\Sigma_E^{-1} \big\{ \dd{^2}{\eta^2_{n,E}} \big[-f(\eta_{n,E})\big]\big\}\Sigma_E^{-1}.
	\end{align*}
	To compute the derivatives of $f$, using \eqref{eq:b_and_gtilde} we note that
	\begin{align*}
		\dd{}{\eta_{n,E}} \left[-f(\eta_{n,E})\right]
		&=-\sqrt{n}\widehat{\beta}_{n,E} + \dd{}{\eta_{n,E}}h^*(\eta_{n,E}, 0_{|E|})
		=-\sqrt{n}\widehat{\beta}_{n,E} + \widetilde{b}^*.
	\end{align*}
	Thus,
	$-\nabla^2f(\eta_{n,E}) = \dd{\widetilde{b}^*}{\eta_{n,E}}.$
	To calculate this Hessian, we start from \eqref{eq:b_and_gtilde},
	which gives us the following equalities
	\begin{align}
	\label{h'1}
		\bar{\Theta}^{-1}\widetilde{b}^* - \bar{A}^\top\bar{\Omega}^{-1}\left( \widetilde{g}^* - \bar{A} \widetilde{b}^* - \bar{b}\right) &= \eta_{n,E}\\
		\label{h'2}
		\bar{\Omega}^{-1} (\widetilde{g}^* - \bar{A} \widetilde{b}^* - \bar{b})
		-\nabla \log J_{\Pi}(\widetilde{g}^*, u_n)
		+ \nabla{\text{\normalfont Barr}}_{\mathcal{K}}(\widetilde{g}^*) &= 0_{|E|},
	\end{align}
where $\nabla \log J_{\Pi}(\widetilde{g}^*, u_n) = \frac{\partial}{\partial g}\log J_{\Pi}(g, u_n)|_{\widetilde{g}^*}$.
	Differentiating \eqref{h'1} with respect to $\eta_{n,E}$,
	\begin{align*}
		\left(\bar{\Theta}^{-1} + \bar{A}^\top \bar{\Omega}^{-1} \bar{A} \right)\dd{\widetilde{b}^*}{\eta_{n,E}}
		-
		\bar{A}^\top \bar{\Omega}^{-1} \dd{\widetilde{g}^*}{\widetilde{b}^*} \dd{\widetilde{b}^*}{\eta_{n,E}} &= I_{|E|}
	\end{align*}
	and therefore
	\begin{align*}
		\dd{\widetilde{b}^*}{\eta_{n,E}} = \left(\bar{\Theta}^{-1} + \bar{A}^\top \bar{\Omega}^{-1} \bar{A}-\bar{A}^\top \bar{\Omega}^{-1} \dd{\widetilde{g}^*}{\widetilde{b}^*} \right)^{-1}.
	\end{align*}
	Differentiating equation \eqref{h'2} with respect to $\widetilde{b}^*$, we have
	\begin{align*}
		\bar{\Omega}^{-1}\dd{\widetilde{g}^*}{\widetilde{b}^*}
		-
		\bar{\Omega}^{-1}\bar{A}
		-
		\nabla^2 \log J_{\Pi}(\widetilde{g}^*, u_n)\dd{\widetilde{g}^*}{\widetilde{b}^*}
		+
		\nabla^2{\text{\normalfont Barr}}_{\mathcal{K}}(\widetilde{g}^*)\dd{\widetilde{g}^*}{\widetilde{b}^*} &= 0_{|E|}
	\end{align*}
	and therefore
$ 
		\partial{\widetilde{g}^*}/\partial{\widetilde{b}^*} =
		\left( \bar{\Omega}^{-1}
		- \nabla^2 \log J_{\Pi}(\widetilde{g}^*, u_n)
		+ \nabla^2{\text{\normalfont Barr}}_{\mathcal{K}}(\widetilde{g}^*)
		\right)^{-1} \bar{\Omega}^{-1}\bar{A}.
$ 
From \eqref{eq:b_and_gtilde}, \eqref{eq:etaprime_u_nablah} and taking $\eta_{n,E} = \eta_{n,E}^*$,
it follows that $\widetilde{b}^*=\sqrt{n}\widehat{\beta}_{n,E}$ and $\widetilde{g}^*=g_n^*(\sqrt{n}\widehat{\beta}_{n,E})$.
		It follows that $-\nabla^2f(\eta_{n,E}^*)$ 
is equal to $M^{-1}$.
\end{proof}

\subsubsection{Proof of Lemma 4.2}

\begin{proof}
First, let the matrix
 $
 Q=\nabla^2{\text{\normalfont Barr}}_{\mathcal{K}}(g_n^*(\sqrt{n}\widehat{\beta}_{n,E})) - \nabla^2 \log J(g_n^*(\sqrt{n}\widehat{\beta}_{n,E}); U_n),
 $
 which is positive definite.
 Using the expressions of $\bar \Theta$, $\bar\Omega$ and $\bar A$, we can rewrite the matrix $M$ from Theorem 3.1 as
 \begin{equation*}
  M = \Sigma_E^{-1} +  A_{\mathcal{E}}^T \Omega^{-1/2} (I + \Omega^{-1/2} B_{\mathcal{E}}\widehat{U}_nQ^{-1} (B_{\mathcal{E}}\widehat{U}_n)^\top\Omega^{-1/2})^{-1}\Omega^{-1/2} A_{\mathcal{E}}.
  \end{equation*}
 Since for any symmetric positive definite matrix $A$ with elements $a_{ij}$ it holds that $\max_{i,j}|a_{ij}|\le \lambda_{\max}(A)$, we look for an upper bound
 on the largest eigenvalue of $I^{-1}_{n,\text{mle}}= n^{-1} \Sigma_E M\Sigma_E$.
 We use the property \citep[e.g.,][Th.~8.1.5]{GolubVanLoan1996} that for two symmetric matrices $A$ and $B$,
$ \lambda_{\max}(A)+\lambda_{\min}(B) \le \lambda_{\max}(A+B) \le \lambda_{\max}(A)+\lambda_{\max}(B)$ to find first that
 the eigen-values of
 $(I + \Omega^{-1/2} B_{\mathcal{E}}\widehat{U}_nQ^{-1} (B_{\mathcal{E}}\widehat{U}_n)^\top\Omega^{-1/2})^{-1}$
 are upper bounded by $1$,
 and next that
\begin{eqnarray*}
\lambda_{\max}(\Sigma_E M\Sigma_E)
&= & \lambda_{\text{max}}(\Sigma_E + \Sigma_E A_{\mathcal{E}}^\top \Omega^{-1/2} (I + \Omega^{-1/2} B_{\mathcal{E}}\widehat{U}_nQ^{-1} (B_{\mathcal{E}}\widehat{U}_n)^\top\Omega^{-1/2})^{-1}\Omega^{-1/2} A_{\mathcal{E}}\Sigma_E)\\
 &\leq & \lambda_{\text{max}}(\Sigma_E) (1+ \lambda_{\text{max}}(\Sigma_E) \lambda_{\text{max}}(A_{\mathcal{E}}^\top \Omega^{-1} A_{\mathcal{E}})).
\end{eqnarray*}
Using the definition of $u_0$, this proves Lemma 3.2.
\end{proof}

\section{Randomization Covariance}
In the context of data splitting, with the non-empty $S\subset [n]$, denote by $X^S$ and $Y^S$ the submatrices of $X$ and $Y$ consisting of the rows indexed by $S$, where
$n_1= |S|<n$. Define $r= n_1/n$.
Suppose that we obtain the group lasso estimator on the $S$-subset of the data,
\begin{equation*}
\widehat\beta_{n,(S)}^{(\Lambda)} \in \underset{\beta\in\mathbbm{R}^p}{\text{argmin}}\left\{ \frac{1}{\sqrt{n_1}} \ell_S (X^S\beta;Y^S) + \sum_{g\in\mathcal{G}}\lambda_{g, (S)}\|\beta_g\|_2 \right\},
\end{equation*}
with $\ell_S(X^S\beta;Y^S)=\sum_{i=1}^{n_1}\rho((x_i^S)^\top\beta;y_i^S)$ and $\lambda_{g, (S)} =\sqrt{r}\lambda_g$, where $\lambda_g$ are the tuning parameters for our randomized regularized problem in (2) in the article.
If we divide the objective function in the above-stated problem by $\sqrt{r}$, then we note that the regularized problem on the $S$-subset of data can be rewritten as
\begin{equation*}
\widehat\beta_{n,(S)}^{(\Lambda)} \in \underset{\beta\in\mathbbm{R}^p}{\text{argmin}}\left\{ \frac{1}{r\sqrt{n}} \ell_S (X^S\beta;Y^S) + \sum_{g\in\mathcal{G}} \lambda_g\|\beta_g\|_2 \right\}.
\end{equation*}
At the solution, we have, with $\widehat{u}_{n,(S)}$ and $\widehat{z}_{n,(S)}$ now for the $S$-subsample, corresponding to a set of selected indices $E_S$,
 \begin{equation*}
	\begin{aligned}
		0_p &= \frac{1}{r\sqrt{n}} (X^S)^\top \nabla \ell_S(X^S\widehat\beta_{n,(S)}^{(\Lambda)};Y^S) + C_{\mathcal{E}_S}\Lambda_{E_S'} \widehat{z}_{n,(S)} + D_{\mathcal{E}_S}\widehat{u}_{n,(S)}.
	\end{aligned}
\end{equation*}
From the proof of Proposition 2.1, we repeat \eqref{KKT:stationarity},
 \begin{equation*}
	\begin{aligned}
\sqrt{n}\omega_{n} &= \frac{1}{\sqrt{n}} X^\top \nabla \ell (X\widehat\beta_{n}^{(\Lambda)};Y) + C_{\mathcal{E}}\Lambda_{E'} \widehat{z}_n + D_{\mathcal{E}}\widehat{u}_n.
         \end{aligned}
\end{equation*}
This serves as inspiration to define the random variable
\begin{equation}
\widetilde\omega_n = \frac{1}{n} X^\top \nabla \ell (X\widehat\beta_{n,(S)}^{(\Lambda)};Y) - \frac{1}{n_1} (X^S)^\top \nabla \ell (X^S\widehat\beta_{n,(S)}^{(\Lambda)};Y^S).
\label{randomization:defn}
\end{equation}
We easily obtain that under the previous assumptions, it converges to a normal random variable for which the variance matrix is a multiple of the matrix $K$.\\\\\\
\begin{lemma} \label{S-lem:samplesplitting} Under the previous assumptions,
$\sqrt{n}\widetilde\omega_n \overset{d}{\rightarrow} N_p(0_p, K(1-r)/r),$
where $\tilde\omega_n$ is defined according to \eqref{randomization:defn}.
\end{lemma}

\begin{proof}
A Taylor expansion of the gradient of the loss functions around $b_{n,E}^*$ yields us the following asymptotic representation
 \begin{equation}
	\begin{aligned}
\sqrt{n}\widetilde\omega_n &=\frac{1}{\sqrt{n}} X^\top \nabla \ell (X_E\widehat\beta_{n,(S),E}^{(\Lambda)};Y) - \frac{1}{r \sqrt{n}} (X^S)^\top \nabla \ell (X_E^S\widehat\beta_{n,(S),E}^{(\Lambda)};Y^S)\\
                             &= \frac{1}{\sqrt{n}} X^\top \nabla \ell (X_E b_{n,E}^*;Y) - \frac{1}{r \sqrt{n}} (X^S)^\top \nabla \ell (X_E^S b_{n,E}^*;Y^S) + o_p(1).
         \end{aligned}
         \label{asymp:rep:randomization}
\end{equation}
Since $X^S$ is a submatrix of $X$ and by definition of the matrix $K$,
it follows from the representation in \eqref{asymp:rep:randomization} that
$\sqrt{n}\widetilde\omega_n \overset{d}{\rightarrow} N_p\left(0, \frac{1-r}{r}K\right).$
\end{proof}

While $\widetilde\omega_n$ is not the same as $\omega_n$, we may choose $\Omega=K(1-r)/r$ with $r=n_1/n$ to compare the inference from the randomized group lasso estimation method to the procedure of sample splitting where $n_1$ observations are used for the estimation and selection and the remaining $n_2$ observations for inference.

\section{Tables}
\begin{table}[h!]
\tiny
\resizebox{\columnwidth}{!}{%
\begin{tabular}{|l|ccc|}
\hline
{\color[HTML]{000000} } &
  \multicolumn{1}{c|}{{\color[HTML]{000000} Post-GL}} &
  \multicolumn{1}{c|}{{\color[HTML]{000000} Data splitting}} &
  \multicolumn{1}{c|}{{\color[HTML]{000000} Naive}} \\ \hline
{\color[HTML]{000000} Intercept} &
  \multicolumn{1}{c|}{{\color[HTML]{FE0000} (-63.625,123.189)}} &
  \multicolumn{1}{c|}{{\color[HTML]{000000} \textbf{(-268.936,-110.831)}}} &
  {\color[HTML]{FE0000} \textbf{(-156.811,-101.196)}} \\ \hline
{\color[HTML]{000000} Age} &
  \multicolumn{1}{c|}{{\color[HTML]{FE0000} (-5.010,19.812)}} &
  \multicolumn{1}{c|}{{\color[HTML]{000000} -}} &
  {\color[HTML]{FE0000} \textbf{(-17.723,-2.104)}} \\ \hline
{\color[HTML]{000000} Family income to poverty ratio} &
  \multicolumn{1}{c|}{{\color[HTML]{000000} \textbf{(-38.347,-21.485)}}} &
  \multicolumn{1}{c|}{{\color[HTML]{000000} \textbf{(-56.332,-5.861)}}} &
  {\color[HTML]{000000} \textbf{(-34.239,-19.734)}} \\ \hline
{\color[HTML]{000000} BMI} &
  \multicolumn{1}{c|}{{\color[HTML]{FE0000} (-8.461,4.505)}} &
  \multicolumn{1}{c|}{{\color[HTML]{000000} \textbf{(4.118,39.989)}}} &
  {\color[HTML]{FE0000} \textbf{(0.122,11.414)}} \\ \hline
{\color[HTML]{000000} \begin{tabular}[c]{@{}l@{}}DBD895: nr of meals not \\ prepared at home last 7 days\end{tabular}} &
  \multicolumn{1}{c|}{{\color[HTML]{000000} (-22.890,3.407)}} &
  \multicolumn{1}{c|}{{\color[HTML]{000000} -}} &
  {\color[HTML]{000000} (-7.443,5.218)} \\ \hline
{\color[HTML]{000000} \begin{tabular}[c]{@{}l@{}}DBD905: nr of ready-to-eat \\ foods in last 30 days\end{tabular}} &
  \multicolumn{1}{c|}{{\color[HTML]{000000} (-1.685,19.532)}} &
  \multicolumn{1}{c|}{{\color[HTML]{000000} (-15.568,18.090)}} &
  {\color[HTML]{000000} (-0.932,10.241)} \\ \hline
{\color[HTML]{000000} \begin{tabular}[c]{@{}l@{}}DBD910: nr of frozen meals/\\ pizza in last 30 days\end{tabular}} &
  \multicolumn{1}{c|}{{\color[HTML]{009901} \textbf{(15.505,25.776)}}} &
  \multicolumn{1}{c|}{{\color[HTML]{009901} (-9.608,27.253)}} &
  {\color[HTML]{000000} \textbf{(6.014,15.435)}} \\ \hline
{\color[HTML]{000000} Race (ref.: Other)} &
  \multicolumn{1}{c|}{{\color[HTML]{000000} }} &
  \multicolumn{1}{c|}{{\color[HTML]{000000} }} &
  {\color[HTML]{000000} } \\ \hline
\multicolumn{1}{|r|}{{\color[HTML]{000000} Mexican american}} &
  \multicolumn{1}{c|}{{\color[HTML]{009901} \textbf{(-176.805,-23.066)}}} &
  \multicolumn{1}{c|}{{\color[HTML]{009901} -}} &
  {\color[HTML]{000000} (-43.446,8.195)} \\ \hline
\multicolumn{1}{|r|}{{\color[HTML]{000000} Other hispanic}} &
  \multicolumn{1}{c|}{{\color[HTML]{000000} (-91.692,74.963)}} &
  \multicolumn{1}{c|}{{\color[HTML]{000000} -}} &
  {\color[HTML]{000000} (-21.76,29.001)} \\ \hline
\multicolumn{1}{|r|}{{\color[HTML]{000000} Non-hispanic white}} &
  \multicolumn{1}{c|}{{\color[HTML]{000000} (-101.202,25.400)}} &
  \multicolumn{1}{c|}{{\color[HTML]{000000} -}} &
  {\color[HTML]{000000} (-19.094,20.527)} \\ \hline
\multicolumn{1}{|r|}{{\color[HTML]{000000} Non-hispanic black}} &
  \multicolumn{1}{c|}{{\color[HTML]{FE0000} (-118.582,9.161)}} &
  \multicolumn{1}{c|}{{\color[HTML]{000000} -}} &
  {\color[HTML]{FE0000} \textbf{(-61.645,-18.412)}} \\ \hline
{\color[HTML]{000000} Civil state (ref.: Never married)} &
  \multicolumn{1}{c|}{{\color[HTML]{000000} }} &
  \multicolumn{1}{c|}{{\color[HTML]{000000} }} &
  {\color[HTML]{000000} } \\ \hline
\multicolumn{1}{|r|}{{\color[HTML]{000000} Married/Living with}} &
  \multicolumn{1}{c|}{{\color[HTML]{FE0000} (-85.661,3.854)}} &
  \multicolumn{1}{c|}{{\color[HTML]{000000} (-10.760,116.616)}} &
  {\color[HTML]{FE0000} \textbf{(-40.487,-7.548)}} \\ \hline
\multicolumn{1}{|r|}{{\color[HTML]{000000} Widowed/Divorced/Separated}} &
  \multicolumn{1}{c|}{{\color[HTML]{000000} (-22.773,94.319)}} &
  \multicolumn{1}{c|}{{\color[HTML]{000000} (-15.205,120.190)}} &
  {\color[HTML]{000000} (-13.64,24.539)} \\ \hline
{\color[HTML]{000000} Healthy diet (ref.: Poor)} &
  \multicolumn{1}{c|}{{\color[HTML]{000000} }} &
  \multicolumn{1}{c|}{{\color[HTML]{000000} }} &
  {\color[HTML]{000000} } \\ \hline
\multicolumn{1}{|r|}{{\color[HTML]{000000} Excellent}} &
  \multicolumn{1}{c|}{{\color[HTML]{009901} \textbf{(-233.841,-37.775)}}} &
  \multicolumn{1}{c|}{{\color[HTML]{009901} (-136.018,38.822)}} &
  {\color[HTML]{000000} \textbf{(-122.073,-53.631)}} \\ \hline
\multicolumn{1}{|r|}{{\color[HTML]{000000} Very good}} &
  \multicolumn{1}{c|}{{\color[HTML]{000000} \textbf{(-234.798,-74.494)}}} &
  \multicolumn{1}{c|}{{\color[HTML]{000000} \textbf{(-302.579,-95.239)}}} &
  {\color[HTML]{000000} \textbf{(-118.457,-70.005)}} \\ \hline
\multicolumn{1}{|r|}{{\color[HTML]{000000} Good}} &
  \multicolumn{1}{c|}{{\color[HTML]{000000} \textbf{(-220.210,-76.543)}}} &
  \multicolumn{1}{c|}{{\color[HTML]{000000} \textbf{(-189.227,-62.704)}}} &
  {\color[HTML]{000000} \textbf{(-95.894,-58.082)}} \\ \hline
\multicolumn{1}{|r|}{{\color[HTML]{000000} Fair}} &
  \multicolumn{1}{c|}{{\color[HTML]{009901} \textbf{(-206.494,-53.891)}}} &
  \multicolumn{1}{c|}{{\color[HTML]{009901} (-111.645,3.662)}} &
  {\color[HTML]{000000} \textbf{(-60.048,-23.021)}} \\ \hline
{\color[HTML]{000000} Male} &
  \multicolumn{1}{c|}{{\color[HTML]{FE0000} (-11.759,15.493)}} &
  \multicolumn{1}{c|}{{\color[HTML]{000000} \textbf{(-110.965,-16.105)}}} &
  {\color[HTML]{FE0000} \textbf{(-61.818,-35.157)}} \\ \hline
{\color[HTML]{000000} Daily 4/5 or more drinks} &
  \multicolumn{1}{c|}{{\color[HTML]{FE0000} (-29.398,1.617)}} &
  \multicolumn{1}{c|}{{\color[HTML]{000000} \textbf{(17.810,124.061)}}} &
  {\color[HTML]{FE0000} \textbf{(43.756,74.193)}} \\ \hline
{\color[HTML]{000000} High blood pressure/Hypertension} &
  \multicolumn{1}{c|}{{\color[HTML]{FE0000} (-8.023,21.734)}} &
  \multicolumn{1}{c|}{{\color[HTML]{000000} \textbf{(15.698,104.643)}}} &
  {\color[HTML]{FE0000} \textbf{(25.042,52.895)}} \\ \hline
{\color[HTML]{000000} Smoked at least 100 cigarettes in life} &
  \multicolumn{1}{c|}{{\color[HTML]{FE0000} (-26.270,2.679)}} &
  \multicolumn{1}{c|}{{\color[HTML]{000000} (-32.538,60.998)}} &
  {\color[HTML]{FE0000} \textbf{(5.834,32.224)}} \\ \hline
{\color[HTML]{000000} Asthma} &
  \multicolumn{1}{c|}{{\color[HTML]{009901} \textbf{(63.956,93.935)}}} &
  \multicolumn{1}{c|}{{\color[HTML]{009901} (-30.584,70.844)}} &
  {\color[HTML]{000000} \textbf{(23.464,51.932)}} \\ \hline
\end{tabular}%
}\caption{\scriptsize{$90\%$ confidence intervals from our post-selection asymptotic likelihood, data splitting and naive inference for $\lambda=0.5$. Bold text indicates 0 is not part of the interval. Cases where significant effects were found through naive inference, but not taking into account the selection are color coded in red. Cases where significant effects were found with our post-selection asymptotic likelihood, but not by data splitting are indicated in green.}}
\label{S-tab:NHANES_CI}
\end{table}

\end{document}